\documentclass[aps,twocolumn,showpacs,superscriptaddress,a4paper,pre,reprint]{revtex4-1}

\usepackage{graphicx,subfigure}
\usepackage{amssymb, amsmath,amssymb,amsfonts}
\usepackage{amsthm,mathrsfs,amsopn}
\usepackage{dcolumn}
\usepackage{bm}
\usepackage{color}
\usepackage[utf8]{inputenc}
\usepackage[normalem]{ulem} 
\usepackage{rotating}
\theoremstyle{plain} 

\newtheorem{theorem}{Theorem}

\newtheorem{remark}[theorem]{Remark}

\newcommand{\be}{\begin{equation}}
\newcommand{\ee}{\end{equation}}
\newcommand{\BE}{\begin{eqnarray}}
\newcommand{\EE}{\end{eqnarray}}
\newcommand{\BM}{\begin{multline}}
\newcommand{\EM}{\end{multline}}

\begin{document}

\title{Nonlinear walkers and efficient exploration of congested networks}

\author{Timoteo Carletti}
\affiliation{naXys, Namur Institute for Complex Systems, University of Namur, Belgium}
\author{Malbor Asllani} 
\affiliation{MACSI, Department of Mathematics and Statistics, University of Limerick, Limerick V94 T9PX, Ireland}
\author{Duccio Fanelli} 
\affiliation{Dipartimento di Fisica e Astronomia, University of Florence,
INFN and CSDC, Via Sansone 1, 50019 Sesto Fiorentino, Florence, Italy}
\author{Vito Latora}
\affiliation{School of Mathematical Sciences, Queen Mary University of
  London, Mile End Road, E1 4NS, London (UK)}
\affiliation{Dipartimento di Fisica ed Astronomia, Universit\`a di Catania and INFN, I-95123 Catania, Italy}
\affiliation{The Alan Turing Institute, The British Library, London NW1 2DB, United Kingdom}

\begin{abstract}
  Random walks are the simplest way to explore or search a graph, and
  have revealed a very useful tool to investigate and characterize the
  structural properties of complex networks from the real world.  For
  instance, they have been used to identify the modules of a given
  network, its most central nodes and paths, or to determine the
  typical times to reach a target.  Although various types of random
  walks whose motion is biased on node properties, such as the degree,
  have been proposed, which are still amenable to analytical solution,
  most if not all of them rely on the assumption of linearity and
  independence of the walkers.
In this work we introduce a novel class of nonlinear stochastic
processes describing a system of interacting random walkers moving
over networks with finite node capacities. The transition
probabilities that rule the motion of the walkers in our model are
modulated by nonlinear functions of the available space at the
destination node, with a bias parameter that allows to tune the
tendency of the walkers to avoid nodes occupied by other
walkers. Firstly, we derive the master equation
governing the dynamics of the system, and we determine  
an analytical expression for the occupation probability of
the walkers at equilibrium in the most general case, and under
different level of network congestions. 
Then, we study different type of
synthetic and real-world networks, presenting numerical and analytical
results for the entropy rate, a proxy for the
network exploration capacities of the walkers. 
We find that, for each
level of the nonlinear bias, there is an optimal crowding that
maximises the entropy rate in a given network topology.
{The analysis suggests that a large fraction of real-world networks are organised in such a way as to favour exploration under congested conditions.}
Our work provides a general and versatile framework to model nonlinear
stochastic processes whose transition probabilities vary in time
depending on the current state of the system.
\end{abstract}

\maketitle
\section{Introduction}
Random walks are basic stochastic processes, which
bear universal interest in light of their widespread and
cross-disciplinary usage.  Since the pioneering work by Pearson and
Rayleigh, back in 1905 \cite{Pearson05,Rayleigh}, the number of
studies invoking the notion of random walker has grown rapidly, to
eventually cover a broad spectrum of applications, from physics to
engineering, via biology and economics.

Random walks have been
thoroughly studied on regular lattices \cite{Hughes95} and, more
recently, on graphs displaying complex topologies
\cite{BLMCH2006,LNR2017,vespignani,Masuda17}.  In the simplest
possible scenario, the walker moves, with a uniform probability, from
a given node $i$ to one of its neighbours $j$.  Alternatively,
when the dynamics takes place on a weighted graph, one can gauge the
probability of performing the move with the weight of the link $(i,j)$
\cite{Cover1991,Meloni08}. Various other classes of random walkers are
however possible on complex networks \cite{Masuda17}.  The walk
can be for instance biased on the topological properties of the nodes 
of the network, such as the node degree or the betweenness.
In~\cite{Gomez-Gardenes2008}, the probability
for a walker to perform a move is modulated by a power law of the degree
of the target node.  Tuning the scaling exponent enables one to steer
the dynamics towards the hubs or favour, at variance, the motion towards
low-degree nodes. Furthermore, when the nodes are also characterised by
endogenous state variables, mirroring congestion or tagging local
deficiencies, these can be considered as a feedback to modify the
motion of individual agents \cite{Manfredi18}.
Metapopulation models of random walkers
which integrate random relocation moves with local interactions
depending on the node occupation probabilities have also been proposed in
\cite{Cencetti18} and employed to extract information on
the architecture of the underlying network.
Mutual interference, as stemming
from the competition for available spatial resources, is unavoidably
present when many walkers are moving at the same time across the
nodes of a given network \cite{Bagnoli11}.
In Ref.~\cite{AsllaniPRL2017} a model of
transport on networks which accounts for the finite carrying capacity of
the nodes has been proposed. In particular, it has been shown that
the equilibrium density (stationary distribution) of crowded walkers saturates for large enough values of the connectivity, while conventional non-interacting
agents have a stationary distribution which depends linearly 
on the nodes degree. 

In this work we introduce and study a novel {and general class of {\em
    nonlinear} Markov chains with transition probabilities that change
  in time depending on the current state of the system. These describe
  the motion of interacting random walkers whose probability to jump
  to a node of a network is a nonlinear function of the number of
  walkers currently at the node.} Such class of {\em nonlinear random
  walkers} provides a {versatile, but at the same time analytically
  treatable, framework to study} the dynamics of active agents that
modulate their motion depending on the level of perceived congestion
on the network.
As a special case, we will study walkers whose probability to move
from node $i$ to a neighbour $j$ scales as a power law of the
occupation density of node $j$, with an exponent $\sigma \geq 0$ that
measures the anti-social behaviour of the walkers, i.e. their tendency
to avoid nodes already occupied by other walkers.
Under this framework we will prove that, for any given
network and each selected value of $\sigma$, there is always 
an optimal value of the network crowding (the total load on
the network), {that maximises the entropy rate, i.e.}  
facilitates the exploration of the network. 
{We will also show that, in many real-world networks, 
the maximal value of the entropy rate is larger than that in 
randomised networks with the same degree distributions.}

\section{The stochastic process}  
Consider a set of interacting
  agents (walkers) moving on an undirected network with $N$
nodes, each endowed with a finite carrying capacity. 
For the sake of simplicity, we assume that all 
the nodes have the same carrying capacity, i.e. each of them can 
simultaneously host a maximum number of agents 
equal to $M$. The
architecture of the network is described in terms of the binary adjacency
matrix $A= \{ a_{ij} \}$, with $a_{ij}=1$ if there is a link 
connecting nodes $i$ and $j$, while 
$a_{ij}=0$ otherwise. At each time $t$,    
the state of the system (our set of walkers) is
specified by the vector $\mathbf{m}=(m_1,\dots,m_N)$, where $0\leq m_i
\le M$ is the number of agents that belong to node $i$, at time $t$.
The total number of walkers in the network is
fixed in time and is a tunable parameter of the model. We can
control it by introducing the {\em average node crowding} $\beta = 1/N
\sum_{i=1}^N m_i / M $.  By definition, $\beta \in(0,1]$ quantifies
  the average node congestion, with $\beta \rightarrow 0$ corresponding to
  the idealised diluted setting. Hence, we can tune the total
    number of walkers in the network, $\beta M N$, by independently
    changing $M$ and $\beta$.
  Agents perform a biased random walk
  hopping between neighbouring nodes, provided there is enough space
  at the arriving destination. Differently from
  Refs.~\cite{Gomez-Gardenes2008}, the motion of the agents is not
  biased on the topological properties of the underlying graph but on the positions of the other agents in the
  network. More specifically, the bias results in two distinct
  contributions, respectively representing the {\em willingness} to
  leave a node $i$, and the {\em attractiveness} of the target node
  $j$. The first component is a function, $f(x_i)$, of the density
  $x_i=m_i/M$ on node $i$. The second term is made to depend on the
  available space $1-x_j=(M-m_j)/M$ at node $j$, as {$g(x_j) \equiv
  \hat{g}(1-x_j)$}. As a natural constraint, we require that $f(x)$
  vanishes at zero, i.e. $f(0)=0$, since no hops can take place from an
  empty node. Further, we assume that $f(\cdot)$ is a
  non-decreasing nonlinear 
  function of $x$, a choice that amounts to modelling
  anti-social reactions of the walkers
  to enhanced crowded conditions, i.e. their
  tendency to avoid nodes already occupied by other walkers. 
  Observe that
  the standard unconstrained random
  walk is eventually recovered when setting $f(x)=x$ and $g(x)=1$, for
  all $x$. The finite carrying capacity signifies that no transition towards node $j$ can take place, if
  $x_j=1$, namely if the arrival node is fully packed.  We hence
  require the self-consistent condition $g(1)\equiv \hat{g}(0)=0$.
  Any possible choice of $f(x)$ and $g(x)$ fulfilling the above
  prescription is in principle possible. Notice that the linear
  model studied in~\cite{AsllaniPRL2017} can be obtained as a
  particular case of our model if we fix $f(x)=x$ and $g(x)=1-x$. 
  On the other hand, adopting nonlinear
  functions for $g(x)$, enables one to reveal a large plethora of
  interesting dynamical features, which reflect different modalities
  of active reaction to perceived crowding conditions, encompassing
  social/antisocial attitudes.

  The evolution of our system of nonlinear interacting random walks  
  is ruled by the master equation: 
\begin{equation*}
\frac{d}{dt}P(\mathbf{m},t)=\sum_{\mathbf{m}'} \left[ T(\mathbf{m}|\mathbf{m}') P(\mathbf{m}',t)  -   T(\mathbf{m}'|\mathbf{m}) P(\mathbf{m},t) \right] \, 
\end{equation*}
where $P(\mathbf{m},t)$ denotes the probability to find the system in
the state $\mathbf{m}$ at time $t$, $T(\mathbf{m}'|\mathbf{m})$ is the 
transition probability from state $\mathbf{m}$ to state $\mathbf{m}'$, 
and the sum is restricted to states
$\mathbf{m}'$ compatible with $\mathbf{m}$ \cite{Fanelli10}. 
Because the transitions
involve pairs $(i,j)$ connected by a link, i.e. such that $a_{ij}=1$
and only increments and decrements by one unity are allowed, we get
$\mathbf{m}'=(\dots,m_i\pm 1,\dots , m_j\mp1,\dots)$.
The transition probabilities read: 
\begin{equation*}
T(m_i-1,m_j+1| m_i,m_j)=\frac{a_{ij}}{k_i} f\left(\frac{m_i}{M}\right) g\left(\frac{m_j }{M}\right)\, ,
\end{equation*}
{where $k_i=\sum_j a_{ij}$ is the degree of node $i$.} {To make the notation compact, in the above expression
  we solely highlight the state components which are modified by the
  occurring transition ~\cite{McKane05,Lugo08,Biancalani10,Asllani13,Anna10}.
  The
  calculation is however exact: all components are accounted for, and no approximation is involved (see Appendix~\ref{ssec:Master2Dens}).}

A straightforward manipulation yields {(see Appendix~\ref{ssec:Master2Dens}
 and~\cite{AsllaniPRL2017,vankampen})} the following equation for the time evolution of the mean-field node density $\rho_i(t)=\lim_{M\rightarrow \infty}\langle m_i\rangle /M$:
\begin{equation}
\frac{d\rho_i}{dt}=\sum_{j}\Delta_{ij}\Big[ f(\rho_j) g\left(\rho_i \right)-\frac{k_j}{k_i}f(\rho_i)g \left(\rho_j\right) \Big]=\mathcal{L}_i(\rho) \, ,
\label{eq:evolaveni5}
\end{equation}
where $\Delta_{ij}=
a_{ij}/k_j - \delta_{ij}$ is the random walk Laplacian and the
nonlinear operator $\mathcal{L}_i(\rho)$ is defined by the rightmost
equality. 
{Notice that the above mean field equation has been
  obtained by neglecting terms which are $1/M$ smaller than the others.
  This is an approximation at $M$ finite, but holds exactly in the limit
  $M \to \infty$, i.e., when $1/M$ corrections vanish \cite{McKane05}.}
From Eq.~\eqref{eq:evolaveni5} it is immediate to conclude that the
mass, namely the total number of walkers, is an invariant of the dynamics. The quantity $\sum_{i=1}^N \rho_i(t)/N$
is hence conserved and equals to the average node congestion $\beta$.

\section{Equilibrium distribution}  
\begin{figure}[ht]   
\centering
\includegraphics[scale=0.25]{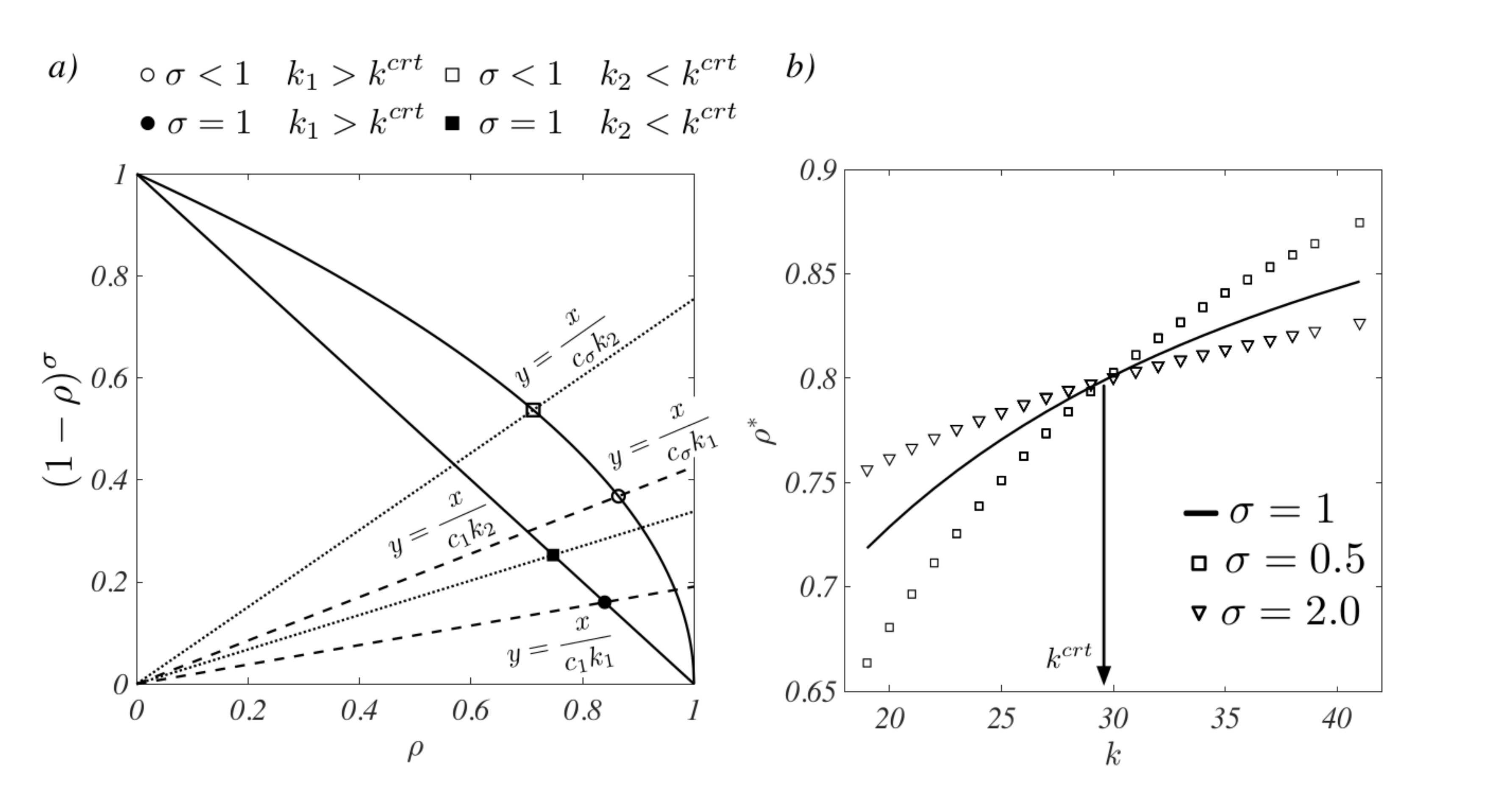}
\vspace{-0.5cm}
\caption{\textbf{The stationary solution}.  (a) The stationary
  distribution of the walkers at a node of degree $k$ is determined as
  the intersection between the line $\rho/(c_{\sigma} k)$ and the
  curve $(1-\rho)^{\sigma}$. The normalisation constant $c_{\sigma}$
  depends on $\sigma$, the level of crowding in the network. (b)
  $\rho^*$ is plotted versus $k$. When $\sigma < 1$ agents cluster on
  nodes with a large degree, while for $\sigma > 1$ hubs are
  progressively depleted (with respect to the case $\sigma=1$). The
  solutions obtained for $\sigma \ne 1$ intersect the curve relative
  to $\sigma=1$ at $k^{\rm{crit}}$.  }
\label{fig:fig1new}
\end{figure}

The stationary solution of
Eq.~\eqref{eq:evolaveni5} can be computed, for any choice of
the nonlinear functions $f$ and $g$ (see Appendix~\ref{ssec:statsol}). We study here the
case in which we set $f(x)=x$ and
$g(x)=(1-x)^\sigma$, with $\sigma \geq 0$.
Modulating the exponent $\sigma$, means selecting 
different exploration strategies of the walkers.
More specifically, the larger $\sigma$ the more the walkers will
  try to avoid densely populated nodes. In other terms, the value of
  $\sigma$ quantifies the level of anti-social behaviour of the
  walkers.
Notice that the diluted limit of non-interacting walkers is 
recovered by letting $\sigma \rightarrow 0$ (and also $\beta
\rightarrow 0$).  
For the case at hand, the stationary solution
$\rho^*_i$ should match the following implicit equation:
\begin{equation}
\frac{\rho^*_i}{c_{\sigma}k_i}={\left(1 - \rho^*_i\right)^{\sigma}}  \quad \forall i\, ,
\label{eq:statsol3bis}
\end{equation}
where $c_{\sigma}$ is a normalisation factor which depends on the
selected $\sigma$. Recalling the definition of $\beta$ yields
$c_{\sigma}=\beta N / \sum_j k_j (1-\rho_j^*)^\sigma$, a condition
which should complement Eq. (\ref{eq:statsol3bis}) for a
self-consistent determination of the stationary equilibrium. To
interpret the above asymptotic solution we will draw a comparison with
that obtained when assuming linear transition rates, $\sigma=1$ When
$\sigma<1$, agents accumulate on the nodes characterised by a large
degree, by consequently depleting those displaying modest connectivity
(see Fig.~\ref{fig:fig1new}). At variance, when $\sigma>1$, hubs are
progressively emptied and the walkers tend to preferentially fill
peripheral nodes with respect to the linear case.
It is instructive to compute the critical degree
$k^{\rm{crit}}$ where such inversion takes place for a 
generic $\sigma$ with respect to the reference case 
$\sigma=1$. A direct computation (see Appendix~\ref{ssec:kcrit}) returns:
\begin{equation}
k^{\rm{crit}}=\left[\left(\frac{c_{1}}{c_{\sigma}}\right)^{1/(1-\sigma)}-1\right]\frac{1}{c_{1}}\, .
\label{eq:kcrit}
\end{equation}
In short, for all $k_i>k^{\rm{crit}}$, we have $\rho_i^*\rvert_{\sigma<1}>\rho_i^*\rvert_{\sigma=1}$, while the opposite inequality holds true if $k_i<k^{\rm{crit}}$. The sign of the inequalities reverse when $\sigma >1$ (see Fig.~\ref{fig:fig1new}).

\section{Exploration under congested conditions} 
The entropy rate of a random walk on a complex network characterises
the walkers ability to explore the network, resulting in a non trivial
indicator where topology and dynamical rules are mutually entangled
\cite{Gomez-Gardenes2008,Burda2009,Sinatra2011}. We will hence
evaluate the entropy rate of the process under study to quantify
the performance of the walkers in exploring a given network under
different level of congestion.  The entropy rate $h$ of a stationary
Markov chain with transition matrix $\Pi= \{\pi_{ij} \}$ and
stationary distribution ${\bf w}^*=\{ w^*_i \}$ can be written as
$h=-\sum_{ij} \pi_{ij} w_i^* \log \pi_{ij}\,$.
In the present case one gets:
\begin{equation}
h =-\sum_{ij}\rho_i^{*} a_{ij}\frac{\rho_i^{*}\left(1-\rho_j^{*}\right)^{\sigma}}{k_i} \log \left[a_{ij}\frac{\rho_i^{*}\left(1-\rho_j^{*}\right)^{\sigma}}{k_i}\right]\, .
\label{eq:entropyrate2}
\end{equation}
The entropy rates depends on the dynamics of the walkers, via the 
stationary probability $\rho_i^{*}$, the nonlinearity exponent
$\sigma$, and the congestion parameter $\beta$, but also on the
structure of the underlying network, via its adjacency matrix $A=\{ a_{ij}
\}$.
The entropy rate in Eq.~\eqref{eq:entropyrate2} (normalised to the system size $N$) can be  
rewritten in the Heterogeneous Mean Field (HMF) approximation, 
by dividing the nodes in different degree classes, 
considering the asymptotic densities of nodes with the  
same degree and performing sums over degree classes (see Appendix~\ref{ssec:hHMF})
~\cite{Gomez-Gardenes2008,HMF1}.

\begin{figure}[ht]  
\centering
\includegraphics[scale=0.21]{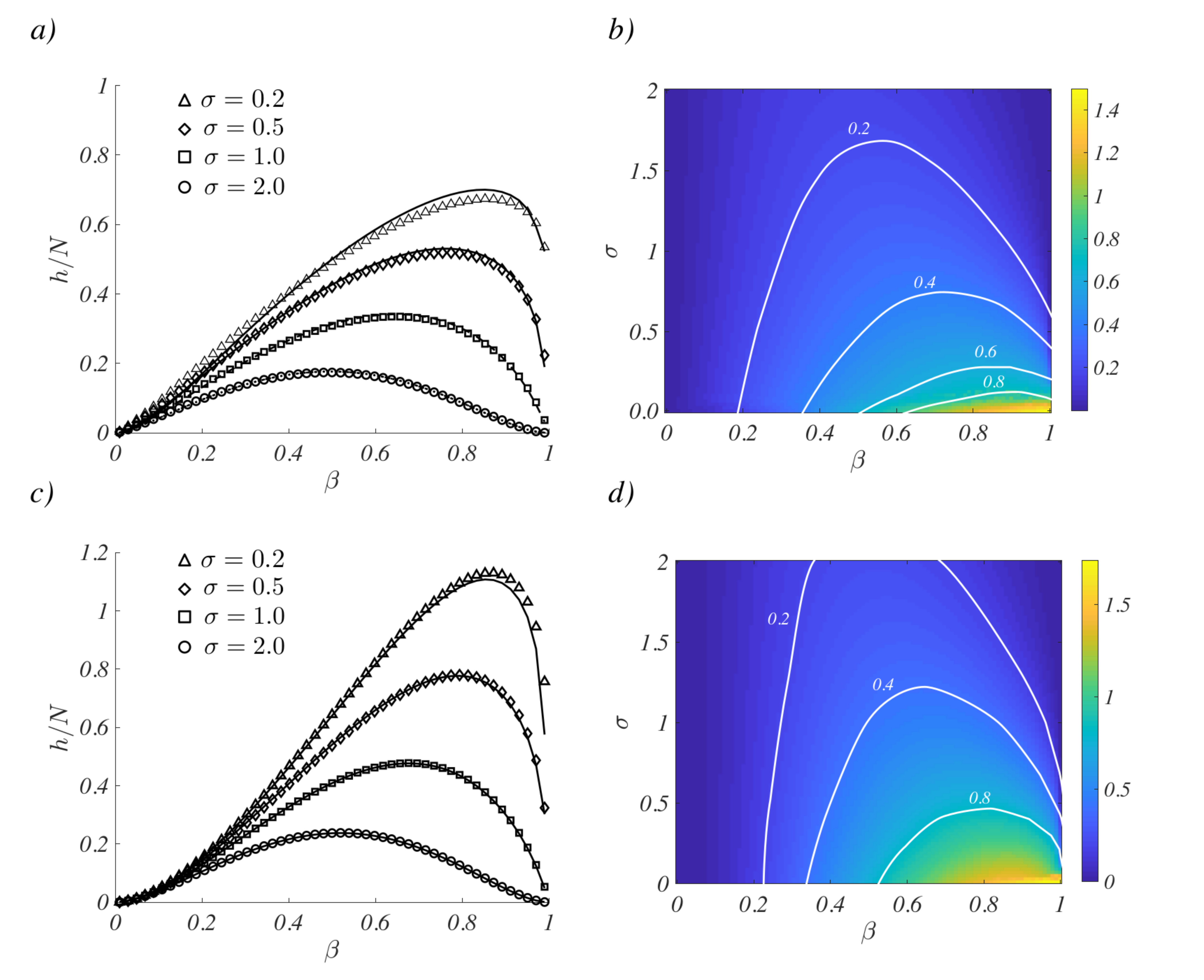}
\vspace{-0.5cm}
\caption{{\bf Entropy rate and iso-{explorability} on
    synthetic networks}. The asymptotic entropy rate per node $h/N$ is
  shown as a function of the average node congestion $\beta$ and for
  different choices of the parameter $\sigma$. Panel a) refers to
  scale-free networks with $N = 1000$, $\gamma= 2.5$ and average
  degree $\langle k \rangle = 6.9$. Panel c) is obtained for Erd\H
  os-{R\'enyi} networks, with $N = 1000$ and $\langle k \rangle = 6.9$.
  Symbols refer to the exact computation performed from
  Eq.~(\ref{eq:entropyrate2}).  Lines are the analytical predictions
  obtained in the {heterogeneous mean field} approximation. In panels b)
  (relative to SF networks) and c) (for the case of Erd\H os-{R\'enyi}
  graphs) the iso-level lines $h/N$ are depicted in the reference
  plane $(\beta, \sigma)$.  A constant level of
  {explorability} is obtained by modulating $\sigma$
  as a nonlinear function of $\beta$.}
\label{fig:entropy}
\end{figure}

Fig.~\ref{fig:entropy} a) and c) show the entropy rate per node,
$h/N$, versus $\beta$, {for synthetic networks with the same
  average degree $\langle k \rangle $ and heterogeneous or homogeneous
  degree distributions respectively.} 
Symbols refer to a direct (and exact)
computation through Eq.~(\ref{eq:entropyrate2}). Solid lines are instead
the results in the HMF approximation.
{We notice that, for any value of $\sigma$, there is 
  an associated value of the crowding parameter $\beta^{\rm opt}$ which
  maximises the entropy rate $h/N$. 
An adequate and network-dependent amount of congestion seems therefore
necessary to favour the network explorability
for any given level of anti-social behaviour (as
measured by the value of $\sigma$). The maximum entropy 
$h^{\rm opt}= h( \beta^{\rm opt})$ increases by decreasing $\sigma$,
  namely when the antisocial behaviour of the 
  walkers is reduced.}
{A trivial global optimum is eventually
obtained when the constraint of a finite carrying capacity is
completely removed.}  
Notice also that the entropy
rate approaches zero when $\beta \rightarrow 1$, namely under
extremely crowded conditions, {i.e. when the agents are practically
  stuck in their positions.} Interestingly, both $\beta^{\rm opt}$ and the value
of $h^{\rm opt}$ depend on the topology
of the network. As an
example, when $\sigma = 0.5$, $\beta^{\rm opt}\sim0.68$ and $h^{\rm
  opt}\sim0.48 \times N$ for Erd\H os-{R\'enyi} random graphs, while
$\beta^{\rm opt}\sim 0.64$
and $h^{\rm opt}\sim0.34 \times N$ for
scale-free networks.
Complementary insights {can be obtained by
  looking at the iso-level lines of $h/N$ in the plane
  ($\beta, \sigma$) reported
  in Fig.~\ref{fig:entropy} b) and d). In order to maintain the
  same level of explorability, the walkers need to adjust the value
  of the dynamic bias $\sigma$, depending on the traffic load $\beta$ in
the network.} 
Intriguingly enough,  $\sigma$ is a non-monotonic function of
$\beta$ on iso-$h$ curves. For small values of $\beta$, 
the walkers have to strengthen their antisocial behaviour (i.e. to
increase $\sigma$) to keep the same value of $h$. Above
a critical value of the average node crowding $\beta$,  
the walkers need instead to weaken their antisocial bias 
(i.e. to decrease $\sigma$).

\begin{figure}[ht]  
\centering
\includegraphics[scale=0.2]{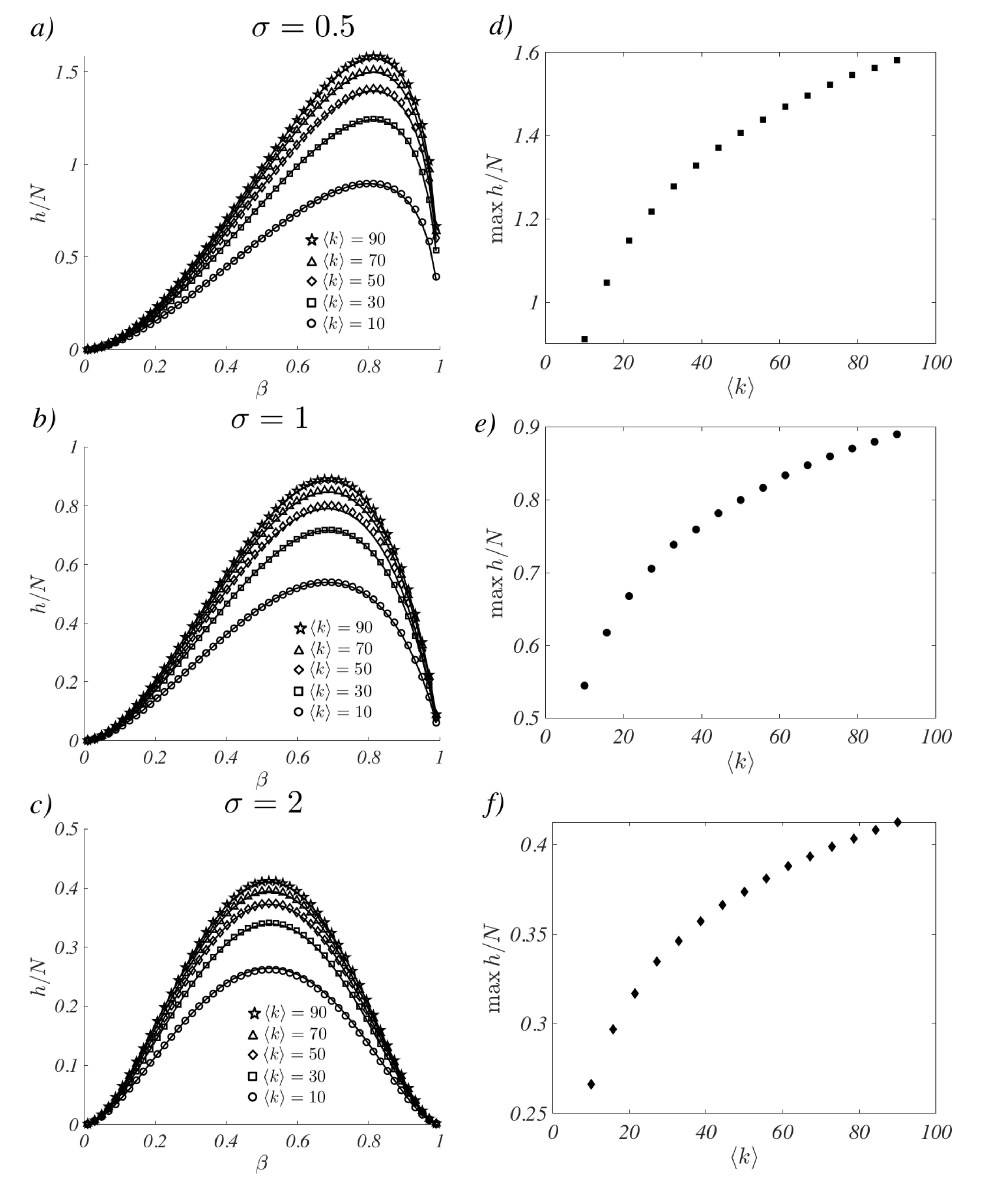}
\caption{\textbf{Entropy rate versus $\langle k \rangle$}. The asymptotic entropy rate $h/N$ per node 
and its maximum value are reported as a function of  the network load (left panels) and the average nodes degree (right panels),  for several values of the nonlinear bias parameter $\sigma$. {Erd\H os-{R\'enyi} networks with $N=100$ nodes have been used.} Lines in the left panels are the analytical 
predictions.}
\label{fig:havek}
\end{figure}

Further, we have analysed how the average node degree of a network,
impacts the entropy rate of the walkers. To this end,
we build different Erd\H os-{R\'enyi} networks with the same number of nodes 
but different average node degrees. Fig.~\ref{fig:havek} shows 
the entropy rate per node as a function of $\beta$ and its
maximum as a function of $\langle k\rangle$, for three values of the
nonlinear bias parameter $\sigma$ ($0.5$ {top} panels, $1.0$ middle
panels and $2.0$ {bottom} panels). Increasing
the network connectivity, yields a global enhancement of the entropy
rate and of its associated maximum. The larger the connectivity, in
fact, the richer the variety of routes available to the motion of the
walkers. 
{As a further point, we stress that random
  architectures return lower entropy values at peak, as compared to
  lattices, a counter-intuitive
  conclusion that is made quantitative in the Appendix~\ref{ssec:kregnet}, where we also derive
  closed analytical formulae for the entropy rate
  on $k$-regular lattices.}

\begin{figure}[ht]
\centering
\includegraphics[scale=0.2]{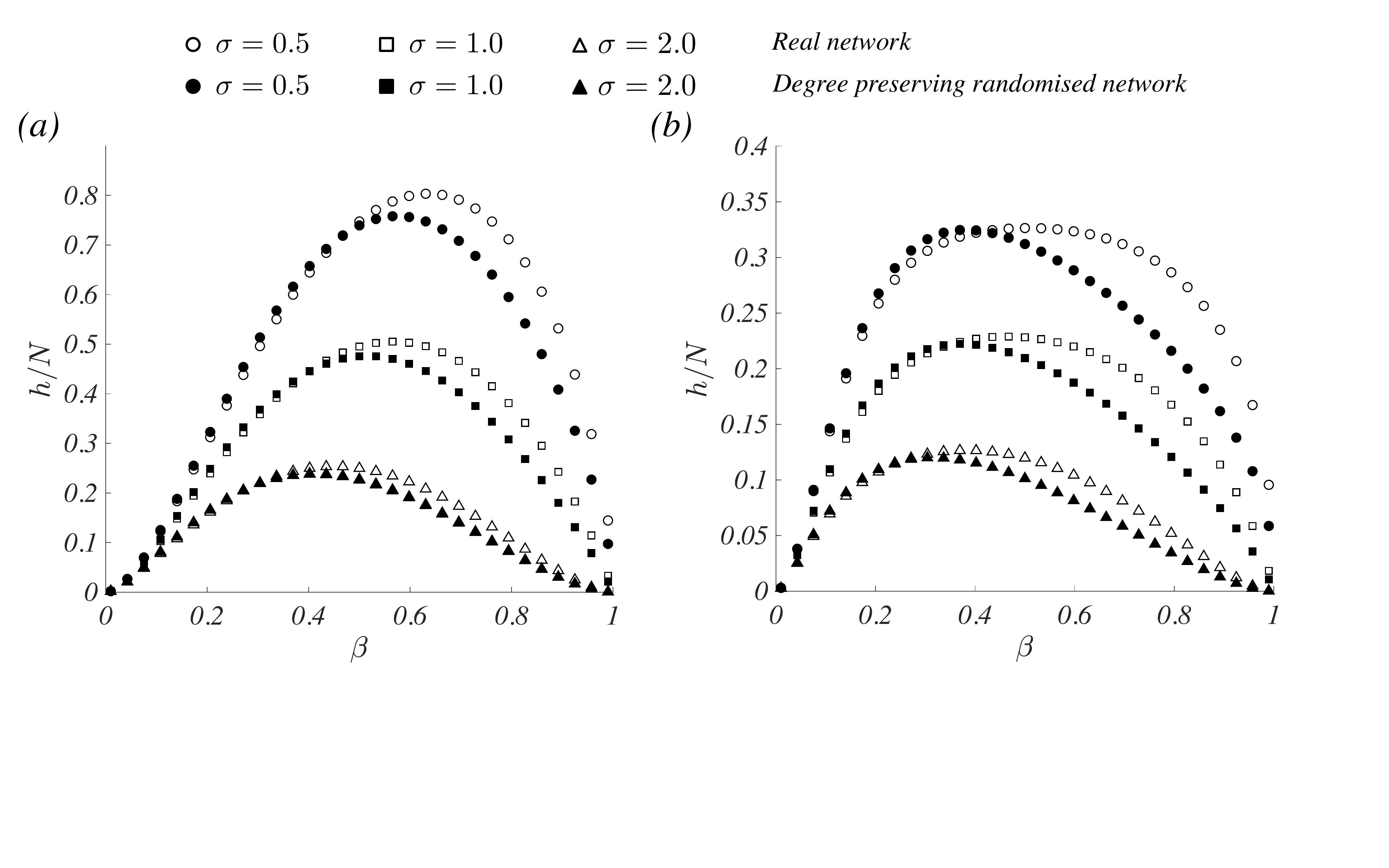}
\vspace{-1.5cm}
\caption{\textbf{Entropy rate for real networks}. The entropy rate per
  node, $h/N$, for different values of $\sigma$ is shown for two real
  networks: (a) the social network of Facebook~\cite{facebook}
  and (b) the transportation network of the $500$ largest US
  airports~\cite{Colizza2007,LNR2017}. {Filled} symbols refer
  to the {average entropy rate obtained for an ensemble of  
  $50$ randomizations which preserve the node degrees of the two real networks.}}
\label{fig:realnet}
\end{figure}

Finally, we studied the properties of our model of nonlinear
  random walkers on several networks taken from the real world. We
computed the entropy rate as a function of the crowding parameter,
determining in each case the optimal values $\beta^{\rm opt}$ and 
$h^{\rm opt}$, for several values of the nonlinear bias
$\sigma$.
%
%
{Results are compared to those obtained on randomized
  versions of the networks. Two different types of
  randomization have been adopted: the first one preserves the degree of
  each node,
  while the second one maintains the network average degree only.}  
As an example Fig.~\ref{fig:realnet} shows the results obtained for: (a) a snapshot of the
social network of Facebook~\cite{facebook}, and (b) for the air transportation
network among the $500$ largest US
airports~\cite{Colizza2007,LNR2017}. First, we confirm the
non-monotonic behaviour of the entropy rate: this latter vanishes for
small and large values of $\beta$, and exhibits a maximum at an
optimal value of the crowding parameter $\beta^{\rm opt}$.
In addition to this, we notice that, for {intermediate and} large
values of $\beta$, the
entropy rate of the walkers on {both these two} real-world networks
is larger than that on the randomised versions of the networks preserving
the degree distribution. 
{In Appendix~\ref{ssec:nets} and Table~\ref{tab:table} we report on the results obtained 
  for a large collection of real networks. Although some of them can also
  exhibit smaller values of entropy rate than their randomised versions, we have
  found that all the networks analysed, which describe urban street patterns, 
  achieve a better explorability.}

\section{Conclusions and Outlook}

Summing up we have here discussed a general approach to the modelling
of biased random walks, under crowded conditions. The formulation of the
problem is not limited to the specific framework analysed here 
(see Appendix~\ref{ssec:morebias} for a generalisation in which also function $f$ is a power law)
and paves the way to devising novel algorithms for an efficient transport
on networks, even in more complex adaptive settings where
the dynamics of the walkers is coevolving with the
underlying network \cite{Iacopini18}.

\section*{Acknowledgements}
{This research used resources of the "Plateforme Technologique de Calcul Intensif (PTCI)" 
(http://www.ptci.unamur.be) located at the University of Namur, Belgium, which is supported  by the 
FNRS-FRFC, the Walloon Region, and the University of Namur (Conventions No. 2.5020.11, GEQ U.G006.15, 
1610468 et RW/GEQ2016). The PTCI is member of the "Consortium des \'Equipements de Calcul Intensif  
(C\'ECI)" (http://www.ceci-hpc.be).  V. L. acknowledges support from the Leverhulme Trust Research Fellowship CREATE: the network components of creativity and success.}

\newpage
\onecolumngrid
\appendix

\section{From the master equation to the deterministic density evolution.}
\label{ssec:Master2Dens}

The goal of this section is derive the mean field equations for the densities, namely Eqs. (3) in the main body of the paper,  from the Master Equation:
\begin{equation}
\frac{d}{dt}P(\mathbf{m},t)=\sum_{\mathbf{m}'} \left[ T(\mathbf{m}|\mathbf{m}') P(\mathbf{m}',t)  -   T(\mathbf{m}'|\mathbf{m}) P(\mathbf{m},t) \right] \, ,
\label{eq:masterA}
\end{equation}
where $P(\mathbf{m},t)$ denotes the probability to find the system at time $t$ in the state $\mathbf{m}=(m_1,\dots, m_N)$. Recall that the above sum is restricted to states $\mathbf{m}'$ compatible with $\mathbf{m}$. Because at a given time only a walker can hop from a given node to one of its neighbours, the states $\mathbf{m}'$ take the form $\mathbf{m}_j=(\dots,m_i\pm 1,\dots , m_j\mp1,\dots)$, for all $j$ with $a_{ij}=1$.

Let us introduce the average number of agents in node $i$ at time $t$, $\langle m_i(t)\rangle \equiv \sum_{\mathbf{m}} m_i P(\mathbf{m},t)$, and the densities $\rho_i(t)=\lim_{M \rightarrow \infty}\langle m_i(t)\rangle /M$. Then, by taking the time-derivative of $\langle m_i\rangle $, recalling~\eqref{eq:masterA} and accounting for the subsets of compatible states we get:
\begin{eqnarray*}
\frac{d}{dt}\langle m_i\rangle&=&\sum_{j,\mathbf{m}_j} m_i\Big[ -T(m_i-1,m_j+1| m_i,m_j)P( m_i,m_j,t)+T(m_i,m_j| m_i+1,m_j-1)P( m_i+1,m_j-1,t) \notag\\
&+&\sum_{j,\mathbf{m}_j} m_i\Big[ -T(m_i+1,m_j-1| m_i,m_j)P( m_i,m_j,t) + T(m_i, m_j| m_i-1,m_j+1)P( m_i-1,m_j+1,t)\Big] \, ,
\end{eqnarray*}
or equivalently
\begin{equation*}
\frac{d}{dt}\langle m_i\rangle=\sum_{j}\Big[ -\langle T(m_i-1,m_j+1| m_i,m_j)\rangle +\langle T(m_i+1,m_j-1| m_i,m_j)\rangle \Big] \, .
\end{equation*}

Consider now the transition probabilities. These latter are expressed in terms two functions, $f$ node and $g$ as:
\begin{equation*}
T(m_i-1,m_j+1|m_i,m_j)=\frac{a_{ij}}{k_i}f\left(\frac{m_i}{M}\right){\hat{g}}\left(\frac{M - m_j }{M}\right)\, ,
\end{equation*}
then we eventually  get:
\begin{equation*}
\frac{d}{dt}\langle m_i\rangle=\sum_{j}\Big[ -\frac{a_{ij}}{k_i}\langle f\left(\frac{m_i}{M}\right) {\hat{g}}\left(\frac{M - m_j }{M}\right)\rangle +\frac{a_{ji}}{k_j} \langle f\left(\frac{m_j}{M}\right){\hat{g}}\left( \frac{M - m_i }{M}\right)\rangle \Big] \, .
\end{equation*}
By introducing the rescale time $\tau=t/M$ and performing the limit $M\rightarrow \infty$ (which in turn amounts to neglecting correlations, i.e. $\langle f(\cdot)\rangle=f(\langle \cdot\rangle)$, similarly for $g$) yields:
\begin{equation*}
\frac{d}{d\tau}\rho_i=\sum_{j}\Big[ -\frac{a_{ij}}{k_i}f(\rho_i) {\hat{g}}\left(1 - \rho_j\right) +\frac{a_{ji}}{k_j} f(\rho_j){\hat{g}} \left(1-\rho_i \right) \Big] \, .
\end{equation*}
By introducing the random walk Laplacian $\Delta_{ij}=a_{ij}/k_j - \delta_{ij}$ and making use of the symmetry of the adjacency matrix, we obtain the sought equation for the time evolution of the density $\rho_i$:
\begin{equation}
\frac{d}{d\tau}\rho_i=\sum_{j}\Delta_{ij}\Big[ f(\rho_j) {\hat{g}}\left(1-\rho_i \right)-\frac{k_j}{k_i}f(\rho_i){\hat{g}} \left(1 - \rho_j\right) \Big]\, .
\label{eq:drhodtau}
\end{equation}

\section{Asymptotic solution}
\label{ssec:statsol}

We now set to calculate the asymptotic density  $\rho_i^*$ as displayed on each node of the network. To do this end we equate to $0$ the right hand side of Eq.~\eqref{eq:drhodtau}, and rewrite the ensuing equation as follows:
\begin{equation*}
\forall i=1,\dots,N \quad 0=\sum_{j}\Delta_{ij}\psi_j(i) \text{ where }\psi_j(i)=\Big[ f(\rho_j) {\hat{g}}\left(1-\rho_i \right)-\frac{k_j}{k_i}f(\rho_i){\hat{g}} \left(1 - \rho_j\right) \Big]\, ,
\end{equation*}
that is for all $i$, $(\psi_1(i),\dots,\psi_N(i))^T$ should be the eigenvector of the random walk Laplace matrix $\mathbf{\Delta}$ associated with the null eigenvalue. In other words, for some constant $\mu(i)$:
\begin{equation*}
\psi_j(i) =k_j \mu(i)\, .
\end{equation*}
Observe that $\psi_i(i)=0$ for all $i$ and thus $k_i \mu(i)=0$, which implies $\mu(i)=0$. Indeed, $k_i\neq 0$ for all $i$ since the network is connected. In conclusion, the asymptotic solution $\rho^*_i$ must satisfy:
\begin{equation*}
f(\rho^*_j) {\hat{g}}\left(1-\rho^*_i \right)-\frac{k_j}{k_i}f(\rho^*_i){\hat{g}} \left(1 - \rho^*_j\right)=0 \quad \forall i,j\, .
\end{equation*}
Reordering the terms one gets:
\begin{equation*}
\frac{f(\rho^*_j)}{k_j {\hat{g}} \left(1 - \rho^*_j\right)} = \frac{f(\rho^*_i)}{k_i {\hat{g}} \left(1 - \rho^*_i\right)}\quad \forall i,j\, .
\end{equation*}
The above condition is met, for any $i$ and $j$, only if the terms on the right and left hand-side equate to a constant $c$ (namely if they do not bear a reflex of the associated index):

\begin{equation}
\frac{f(\rho^*_i)}{k_i {\hat{g}} \left(1 - \rho^*_i \right)}=c\quad \forall i\, .
\label{eq:eqsol}
\end{equation}

For generic $f$ and ${\hat{g}}$ the previous equation can exhibit multiple solutions. To rule out  such possibility, and eventually focus on the interesting setting where just one solution is allowed for, we can assume: (i) $f$ to be non-decreasing function, vanishing at $x=0$; (ii) ${\hat{g}}$ to be non-increasing function, vanishing at $x=1$. In such a way, by continuity, the curves $f(\rho)$ and $ck {\hat{g}}(1-\rho)$ intersect only once, for any choice of $c>0$ and $k>0$.

\subsection{About $k_{crit}$.}
\label{ssec:kcrit}

In Fig.1 (main text) we have shown the non trivial behaviour of the stationary solution $\rho_i^*$ as a function of $\sigma$ and the node degree $k_i$ responsible for the interesting phenomenon of accumulation / depletion of hubs and leaves with respect to the case $\sigma=1$. For any given $\sigma>0$ there exists a unique critical values for the node degree, $k_{crit}$, where such inversion takes place which indirectly defines ``large'' versus ``small'' degrees.

To compute such critical value we need to impose the equality among the stationary solution $\rho^*\rvert_\sigma$, for $\sigma\neq 1$, and the same quantity for $\sigma=1$, $\rho^*\rvert_{\sigma=1}$, both associated to a node with degree $k$. From Eq. (4) (main text) with $\sigma=1$ we obtain $\rho^*\rvert_{\sigma=1}=c_1k/(1+c_1k)$; assuming $\rho^*\rvert_{\sigma=1}=\rho^*\rvert_\sigma$ and substituting this value again in (4) we get
\begin{equation*}
\frac{c_1k_{crit}}{1+c_1k_{crit}}=c_{\sigma}k_{crit}\left(1-\frac{c_1k_{crit}}{1+c_1k_{crit}}\right)^\sigma\,,
\end{equation*}
from which we straightforward obtain
\begin{equation*}
\frac{c_1}{c_{\sigma}}=\frac{1}{\left(1+c_1k_{crit}\right)^{(\sigma-1)}}\,,
\end{equation*}
which gives the Eq. (5) (main text).

{
\subsection{The case of $k$-regular networks}
\label{ssec:kregnet}
The asymptotic solution Eq.~\eqref{eq:eqsol} simplifies in the case of $k$-regular networks, for which i.e. $k_i=k$ for all $i$; in this case indeed, the dependence on the node index $i$ disappears and thus all the nodes will display the same asymptotic density. The total mass conservation allows to determine the latter as
\begin{equation}
\label{eq:kregsol}
\rho_i^*=\beta \quad \forall i=1,\dots N\, ,
\end{equation}
independently of the nonlinear functions $f$ and $g$. These latter are instead used in determining the normalising constant $c$ entering in Eq.~\eqref{eq:eqsol} 
\begin{equation}
\label{eq:kregc}
c=\frac{1}{k}\frac{f(\beta)}{{\hat{g}}(1-\beta)}\, .
\end{equation}
Given the exact asymptotic solution one can explicitly compute the entropy rate given by {Eq.~\eqref{eq:entropyrate2}} (in the main text with the choice $f(x)=x$ and $g(x)=(1-x)^\sigma$ or the following Eq.~\eqref{eq:entropyrateSI}). Indeed the sum over the index $j$ allows to simplify $\sum_j a_{ij}$ with the degree $k_i$ at the denominator, while the second sum returns the factor $N$, being the remaining part independent from $i$. One gets therefore:
\begin{equation*}
{h=-N\beta {f(\beta)g\left(\beta\right)} \log \left[\frac{f(\beta)g\left(\beta\right)}{k}\right]}\, ,
\end{equation*}
{where we have used that $g(x)=\hat{g}(1-x)$.}
}

{Assuming $f(x)=x$ and $g(x)=(1-x)^\sigma$ one can compute the value of $\beta$ which maximises $h$ for a fixed $\sigma$, namely $\beta^{\rm opt}$. Moreover one can calculate the parameter $\sigma^{\rm opt}$ which returns the maximum of $h$ for a fixed $\beta$. To this end one needs to perform the partial derivative, $\partial_{\beta}h$, respectively $\partial_{\sigma}h$, and equating these latter to $0$. In this way one can can draw an interesting conclusion on $\sigma^{\rm opt}$; indeed one can obtain
\begin{equation}
\label{eq:sigmaopt}
\sigma^{\rm opt}=\frac{\log [{k}/{(e\beta)}]}{\log (1-\beta)}\, ,
\end{equation}
and thus if $k<e\beta$ one gets $\sigma^{\rm opt}>0$ while on the contrary one obtains $\sigma^{\rm opt}<0$. The first constraint can be realised only with $k=2$, that is for a $1D$ ring where each node is connected to its two neighbours, one on the left and one on the right, and for $\beta$ sufficiently large, i.e. $\beta >2/e~\sim 0.736$. These facts can explain why in the case of the Erd\H os-{R\'enyi} and scale-free networks one always found $\sigma^{\rm opt}<0$ and thus $h$ is a decreasing function of $\sigma$ (see Fig.~\ref{fig:optbeta}).}

{The computation for $\beta^{\rm opt}$ follows the same reasoning. One can in particular obtain an implicit equation for the optimal value of $\beta$ for a generic function $g(x)$:}
\begin{equation*}
 {\log \frac{\beta^{\rm opt}g(\beta^{\rm opt})}{k}=-1+\frac{g(\beta^{\rm opt})}{g(\beta^{\rm opt})+\left(\beta^{\rm opt}g(\beta^{\rm opt})\right)^\prime}\, .}
\end{equation*}
{For the particular choice $g(x)=(1-x)^\sigma$ we obtain:}
\begin{equation*}
 {\log \frac{\beta^{\rm opt}(1-\beta^{\rm opt})^\sigma}{k}=-1+\frac{1}{2-\frac{\sigma \beta^{\rm opt}}{1-\beta^{\rm opt}}}\, .}
\end{equation*}

\begin{figure}[ht]
\centering
\includegraphics[scale=0.33]{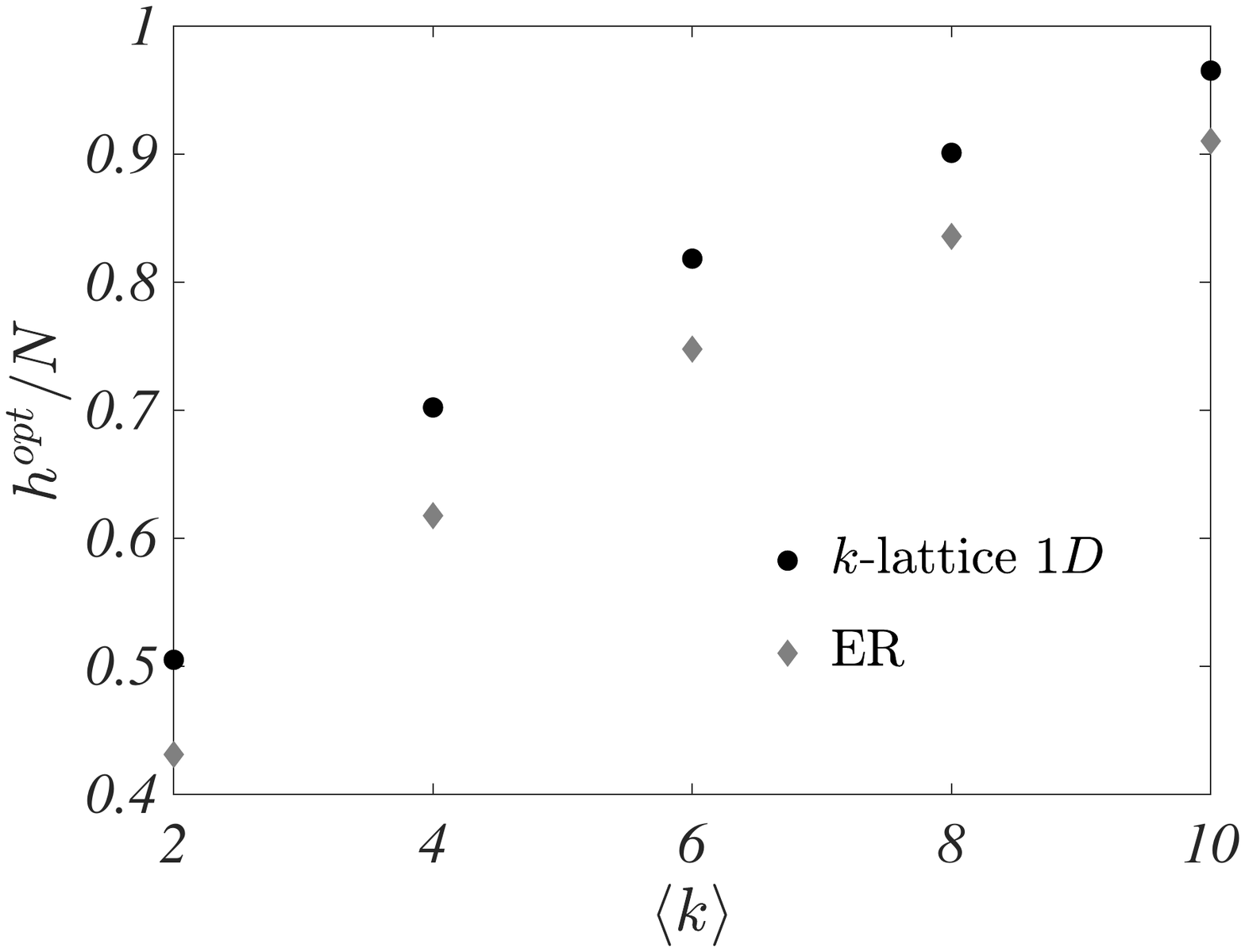}
\vspace{-1.5cm}
\caption{{\textbf{Entropy and topology regularity}. We compare, as a function of the network connectivity, the maximum of the entropy rate per node computed for a $k$-regular $1D$-lattice and an Erd\H os-{R\'enyi} network with the same average degree and the same number of nodes ($N=50$). Results show that the regular topology always exhibits the highest maximum. For large network connectivities the two computed quantities converge to a shared value (indeed both networks converge to the same complete network).}}
\label{fig:hkreger}
\end{figure}

{We conclude this section by observing that more regular topologies can be associated to larger entropy rates and thus to a stronger ergodic behaviour. In particular in Fig.~\ref{fig:hkreger} we compare the maximum of the entropy rate achieved for a $k$-regular $1D$ lattice against the same quantity computed for an Erd\H os-{R\'enyi} network with the same average degree and the same number of nodes. We can observe that for all the values of the average degree, the entropy rate is alway larger in the case of the regular lattice than for the random network.}

{
\begin{remark}[The $k$-Cayley trees]
A similar analysis can be performed in the case of $k$-Cayley trees, where each node has degree $k$ (also called coordination number), but the leaves that by definition have degree $1$. This implies that there will be two values for the asymptotic density, one associated to the leaves, $\rho^*_{out}$, and one for the remaining nodes, i.e. the inner ones, $\rho^*_{inn}$, determined by:
\begin{equation}
\label{eq:cayley}
\frac{\rho^*_{inn}}{k \left(1 - \rho^*_{inn} \right)^\sigma}=c \quad \text{and}\quad \frac{\rho^*_{out}}{\left(1 - \rho^*_{out} \right)^\sigma}=c\, .
\end{equation}
The constraint on the conservation of the total mass and the observation that in the limit of infinitely large Cayley tree, i.e. for a diverging number of shells, the number of inner nodes divided by the number of leaves converges to $1/(k-2)$, provide a third relation:
\begin{equation}
\label{eq:cayley2}
\rho^*_{inn}\frac{1/(k-2)}{1+1/(k-2)}+\rho^*_{out}\frac{1}{1+1/(k-2)}=\beta\, .
\end{equation}
From Eqs.~\eqref{eq:cayley} and~\eqref{eq:cayley2} one can determine the three variables $\rho^*_{inn}$, $\rho^*_{out}$ and $c$, and then again the entropy rate $h$. Let us observe that in this limiting case the average degree of the Cayley tree converges to $2$ and thus we cannot fairly compare its entropy rate with the one obtained for the $k$-regular $1D$ lattice or the Erd\H os-{R\'enyi} network with the same average degree.
\end{remark}
}

\subsection{Analytical approximation for the asymptotic solution, when $f(x)=x$ and $g(x)=(1-x)^{\sigma}$}
\label{ssec:apprsol}

Assuming $f(x)=x$ and $g(x)=(1-x)^{\sigma}$, $\sigma>0$, for $0\leq x <1$ and $0$ otherwise, the asymptotic solution for the density Eq.~\eqref{eq:eqsol} is implicitly given by
\begin{equation}
\rho^*_i= {k_i c}\left(1 - \rho^*_i\right)^{\sigma}\quad \forall i\, .
\label{eq:asymsol}
\end{equation}
In the following we shall write $\rho^*_i(\sigma)$ to stress the dependence on the parameter $\sigma$. For $\sigma=1$ the solution to the latter problem takes the form~\cite{AsllaniPRL2017}
\begin{equation*}
\rho^*_i(1) =\frac{k_i c}{1+k_i c}\quad \forall i\, .
\end{equation*}
Let us introduce $y_i=1-\rho_i^*$ and rewrite the equation for the implicit solution as
\begin{equation}
1-y= {\kappa}y^{\sigma}\, ,
\label{eq:impsol}
\end{equation}
where for a sake of clarity we dropped the index $i$ and we introduced $\kappa=k_ic$. 
One can thus look for a series expansion of $y(\sigma)$ in terms of $(\sigma-1)$, that should converge in a neighbourhood of
$\sigma=1$:
\begin{equation*}
y(\sigma)  = \sum_{n\geq0} \frac{y_n}{n!}(\sigma-1)^n\, .
\end{equation*}
Inserting this power series into Eq.~\eqref{eq:impsol}, recalling that 
\begin{equation*}
\frac{d}{d\sigma}[y(\sigma)]^{\sigma}=[y(\sigma)]^{\sigma}\log y+ \sigma[y(\sigma)]^{\sigma-1}\frac{dy}{d \sigma}\, ,
\end{equation*}
and equating terms corresponding to the same powers of $(\sigma-1)$ on the left and the right hand sides of Eq.~\eqref{eq:impsol}, 
we can express $y_n$ as a function of the terms $y_l$, $0\leq l\leq n-1$. This recursive (infinite) system of equations can be explicitly solved. The first few terms are given by
\begin{eqnarray*}
y_0&=&\frac{1}{1+\kappa}\\ 
y_1&=&-\frac{\kappa}{1+\kappa}y_0\log y_0 \\
y_2&=&-\frac{\kappa}{1+\kappa}\left[\left(y_0\log y_0+y_1\right)\log y_0+y_1\right]+\frac{\kappa^2}{(1+\kappa)^2}y_0\log y_0(1+\log y_0)\, .
\end{eqnarray*}
Back to $\rho_i^*(\sigma)$ we obtain
\begin{equation}
\label{eq:analapp}
\rho_i^*(\sigma)=\frac{c k_i}{1+c k_i}+\frac{c k_i}{(1+c k_i)^2}\log (1+c k_i)\times(\sigma-1)+\rho_{i,2}^{*} \times\frac{(\sigma-1)^2}{2}+\mathcal{O}[(\sigma-1)^3]\, ,
\end{equation}
with 
\begin{equation*}
\rho_{i,2}^{*}=\frac{ck_i\log (1+ck_i)}{(1+ck_i)^2}\left[ \left(\log (1+ck_i)-\frac{ck_i}{(1+ck_i)}\log (1+ck_i)\right)
-\frac{ck_i}{(1+ck_i)}\right]-\frac{(ck_i)^2}{(1+ck_i)^2}\frac{\log (1+ck_i)}{1+ck_i}\left[1-\log(1+ck_i)\right]\, .
\end{equation*}

We can thus write, for $\sigma\sim 1$, the following approximate solution: 
\begin{equation}
\rho_i^{*}(\sigma)  = \rho_{i,0}^{*}+\rho_{i,1}^{*}\times (\sigma-1)+\rho_{i,2}^{*} \times \frac{(\sigma-1)^2}{2}+\mathcal{O}{(\sigma-1)^3}\, .
\label{eq:analsol}
\end{equation}
From the explicit form of the coefficients $\rho_{i,n}^*$ one can analyse the dependence of the asymptotic density on the nodes degree and on the normalising parameter $c$, that, we recall, is a function of the crowding amount $\beta$, for fixed $\sigma$. Indeed we observe that for $ck_i\gg 1$ the zeroth order correction is of the order of the unity, $\rho_{i,0}^{*}\rightarrow 1$, while the high order corrections, $n\geq 1$, do satisfy $\rho_{i,n}^{*}=\mathcal{O}{\left((\log ck_i)^n/ck_i\right)}$. Thus, they are negligible provided at least one among $k_i$ and $c$ is sufficiently large. The former condition implies that $i$ is a hub, the latter amounts to operate under crowded conditions, namely $\beta\rightarrow 1$, which in turn implies $c\rightarrow \infty$.
On the other hand,  $ck_i\ll1$ (the network is connected and thus $k_i\neq 0$ for all $i$) yields $\rho_{i,n}^*=\mathcal{O}((ck_i)^{n+1})$; hence, in very diluted conditions, $\beta\rightarrow 0$,  high degree nodes can exhibit a very low density.

\begin{figure*}[ht]
\centering
\includegraphics[scale=0.23]{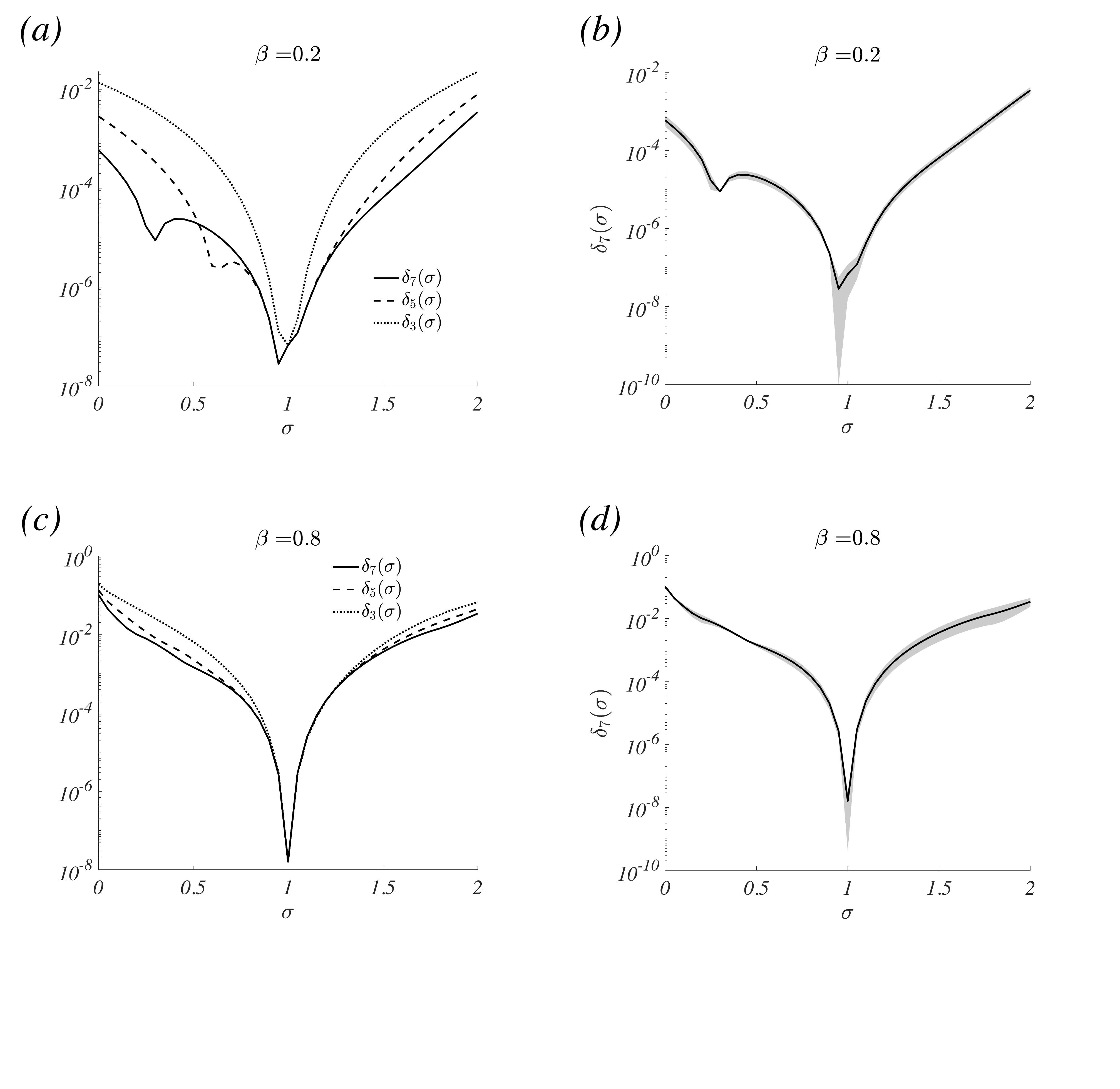}
\vspace{-1.5cm}
\caption{\textbf{Analytical approximation for the asymptotic solution}. Left panels (a \& c): We compare the approximate formula up to the $m$-th order ($m=3$ dotted line, $m=5$ dashed line and $m=7$ solid line) with the exact solution of Eq.~\eqref{eq:asymsol}, as obtained via numerical methods. Right panels (b \& d): for a fixed approximation order, $m=7$, we compute the average error $\delta_7(\sigma)$ over $10$ realisations of the underlying network. The boundaries of the grey shadow are at one standard deviation from the mean. Top panels refer to $\beta=0.2$ while bottom ones to $\beta=0.8$. The underlying network is generated according to the  Erd\H os-{R\'enyi} recipe, with $N=100$ nodes and probability for the existence of a link $p=0.2$.}
\label{fig:analapp}
\end{figure*}

To check the accuracy of the approximation we quantify the discrepancy between the approximate formula Eq.~\eqref{eq:analapp}, up to a given order $m$, $\rho_{i}^{(m)}(\sigma)$, and the exact numerical solution of Eq.~\eqref{eq:asymsol}, $\rho_{i}^{*}(\sigma)$, both for the same fixed value of $\sigma$. The error is specifically defined as:
\begin{eqnarray*}
  \delta_m(\sigma) = \max_{i=1,\dots,N}\Big| 1- \frac{\rho_{i}^{(m)}(\sigma)}{\rho_{i}^{*}(
    \sigma)}\Big|\, .
\end{eqnarray*}
Results reported in Fig.~\ref{fig:analapp} testify on the accuracy of the proposed approximation for $\sigma$ close to $1$; observe that for over a significant window in $\sigma$, the error stays bounded to a  few percents. The actual error depends also on the crowding parameter $\beta$ and on the network topology. The top panels of Fig.~\ref{fig:analapp} refer to a weakly crowded environment, $\beta=0.2$, while bottom ones are obtained when considering a more pronounced degree of imposed crowding, $\beta=0.8$. One can observe that the error deteriorates, as $\beta$ increases. To test the impact of the topology of the underlying network, we created $10$ Erd\H os-{R\'enyi} networks made by $N=100$ nodes and assuming a probability for the existence of a link $p=0.2$. For each network, we computed $\delta_7(\sigma)$. The solid line in the right panels of Fig.~\ref{fig:analapp} displays the average of the computed errors, while the boundaries of the grey shadow are set at one standard deviation by the mean. 
A similar behaviour (data not shown) is obtained when employing different schemes of network generation (as adopted in the main body of the paper). 

\section{Entropy rate and the Heterogeneous Mean Field hypothesis.}
\label{ssec:hHMF}

The aim of this section is provide additional information on the  application of the approximate Heterogeneous Mean Field hypothesis (HMF). Working under this assumption, we will characterise the entropy rate and then derive a simplified formula which holds when correlations among nodes degree can be neglected.

Consider again the entropy rate given by:
\begin{equation}
\label{eq:entropyrateSI}
h=-\sum_{ij}\rho_i^{*} a_{ij}\frac{f(\rho_i^{*})g\left(1-\rho_j^{*}\right)}{k_i} \log \left[a_{ij}\frac{f(\rho_i^{*})g\left(1-\rho_j^{*}\right)}{k_i}\right]\, ,
\end{equation}
where $a_{ij}$ is the adjacency matrix of the underlying network, $(k_i)_{1\leq 1\leq N}$ the nodes degree and  $\rho_i^{*}$ the stationary probability. The first step consists in reorganising the sums as follows
\begin{equation*}
h=-\sum_{i}\rho_i^{*} \frac{f(\rho_i^{*})}{k_i} \log \frac{f(\rho_i^{*})}{k_i}\sum_ja_{ij}g\left(1-\rho_j^{*}\right)-\sum_{i}\rho_i^{*}\frac{f(\rho_i^{*})}{k_i}\sum_ja_{ij}g\left(1-\rho_j^{*}\right)\log \left(a_{ij}g\left(1-\rho_j^{*}\right)\right)\, .
\end{equation*}
Then we invoke the Heterogeneous Mean Field hypothesis, namely we aggregate together nodes which share the same connectivity. Instead of summing on the node's index, we perform the sum on the degree~\cite{Gomez-Gardenes2008}. Let thus denote by $P(k)$ the probability for a generic node to have degree $k$ and let $P(k'|k)$ the conditional probability that a generic node with degree $k$ is connected to a node with degree $k'$, then :
\begin{equation*}
h_{\rm HMF}=\frac{h}{N}=-\sum_{k}P(k)\hat{\rho}_k^{*} \frac{f(\hat{\rho}^{*}_k)}{k} \log \frac{f(\hat{\rho}^{*}_k)}{k}\sum_{k'} P(k'|k)k g\left(1-\hat{\rho}^{*}_{k'}\right)-\sum_{k}P(k)\hat{\rho}^{*}_k\frac{f(\hat{\rho}^{*}_k)}{k}\sum_{k'}P(k'|k)k g\left(1-\hat{\rho}^{*}_{k'}\right)\log \left(g\left(1-\hat{\rho}^{*}_{k'}\right)\right)\, ,
\end{equation*}
where $\hat{\rho}_k^*$ is the density of the nodes that share connectivity $k$.

Assuming an uncorrelated network, $P(k'|k)=k'P(k')/\langle k\rangle$, we get: 
\begin{equation*}
h_{\rm HMF}^{uncorr}=-\sum_{k}P(k)\hat{\rho}_k^{*} f(\hat{\rho}^{*}_k) \log \frac{f(\hat{\rho}^{*}_k)}{k}\sum_{k'} \frac{k'P(k')}{\langle k\rangle} g\left(1-\hat{\rho}^{*}_{k'}\right)-\sum_{k}P(k)\hat{\rho}^{*}_k f(\hat{\rho}^{*}_k)\sum_{k'}\frac{k'P(k')}{\langle k\rangle} g\left(1-\hat{\rho}^{*}_{k'}\right)\log \left(g\left(1-\hat{\rho}^{*}_{k'}\right)\right)\, ,
\end{equation*}
and using the equilibrium definition, $g\left(1-\hat{\rho}_{k'}^{*}\right)=f(\hat{\rho}_{k'}^{*})/(ck')$, we eventually get:
\begin{equation*}
h_{\rm HMF}^{uncorr}=-\sum_{k}P(k)\hat{\rho}_k^{*} f(\hat{\rho}^{*}_k) \log \frac{f(\hat{\rho}^{*}_k)}{k}\sum_{k'} \frac{P(k')}{\langle k\rangle} \frac{f(\hat{\rho}_{k'}^{*})}{c}-\sum_{k}P(k)\hat{\rho}^{*}_k f(\hat{\rho}^{*}_k)\sum_{k'}\frac{P(k')}{\langle k\rangle} \frac{f(\hat{\rho}_{k'}^{*})}{c}\log \frac{f(\hat{\rho}_{k'}^{*})}{ck'}\, ,
\end{equation*}
and after some straightforward computations
\begin{eqnarray*}
h_{\rm HMF}^{uncorr}&=&-\frac{1}{c\langle k\rangle}\left[ \langle \hat{\rho}_k^{*} f(\hat{\rho}^{*}_k) \log \frac{f(\hat{\rho}^{*}_k)}{k}\rangle \langle f(\hat{\rho}_{k}^{*})\rangle+\langle \hat{\rho}_k f(\hat{\rho}_k)\rangle \langle f(\hat{\rho}_{k}^{*})\log \frac{f(\hat{\rho}_{k}^{*})}{ck}\rangle\right] =\notag\\
&=&-\frac{1}{c\langle k\rangle}\Big[ \langle \hat{\rho}_k^{*} f(\hat{\rho}^{*}_k) \log f(\hat{\rho}^{*}_k)\rangle \langle f(\hat{\rho}_{k}^{*})\rangle - \langle \hat{\rho}_k^{*} f(\hat{\rho^{*}}_k) \log {k}\rangle \langle f(\hat{\rho}_{k}^{*})\rangle\notag\\
&+&\langle \hat{\rho}^{*}_k f(\hat{\rho}^{*}_k)\rangle \langle f(\hat{\rho}_{k}^{*})\log f(\hat{\rho}_{k}^{*})\rangle-\langle \hat{\rho}^{*}_k f(\hat{\rho}^{*}_k)\rangle \langle f(\hat{\rho}_{k}^{*})\log {ck}\rangle\Big]\, .
\end{eqnarray*}

Recalling that $f(x)=x$ (for the case analysed in the main body of the paper) we get:
\begin{equation}
h_{\rm HMF}^{uncorr}=-\frac{1}{c\langle k\rangle}\Big[ \langle (\hat{\rho}_k^{*})^2  \log \hat{\rho}^{*}_k\rangle \langle \hat{\rho}_{k}^{*}\rangle - \langle (\hat{\rho}_k^{*})^2 \log {k}\rangle \langle \hat{\rho}_{k}^{*}\rangle+\langle (\hat{\rho}^{*}_k)^2 \rangle \langle \hat{\rho}_{k}^{*}\log \hat{\rho}_{k}^{*}\rangle-\langle (\hat{\rho}^{*}_k)^2 \rangle \langle \hat{\rho}_{k}^{*}\log {ck}\rangle\Big]\, .
\label{eq:entropyrate9}
\end{equation}

\begin{figure}[ht]  
\centering
\includegraphics[scale=0.30]{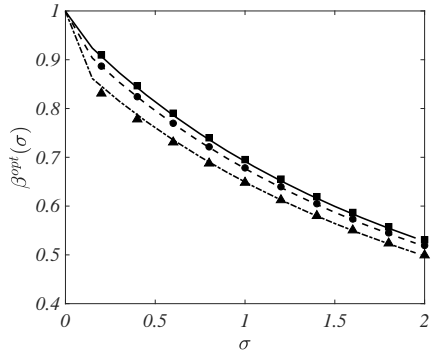}
\vspace{-1.5cm}
\caption{\textbf{Optimal value for the crowding coefficient $\beta^{\rm opt}$ as a function of $\sigma$}. We report
$\beta^{\rm opt}(\sigma)$ as numerically obtained by means of the HMF hypothesis (curves), under the assumption $P(k)\sim 1/k^{\gamma}$ ($\gamma=3$ dashed line, $\gamma=2.5$ dash-dotted line and $\gamma=3.5$ solid line) and compare it with the  corresponding quantity estimated for scale free networks which implement an identical scaling exponent ($\gamma=2.5$ triangles, $\gamma=3.0$ circles and $\gamma=3.5$ squares). Each symbol is the average over $10$ different networks realisations.}
\label{fig:optbeta}
\end{figure}

In the main text we have shown that the entropy rate per node is a
non-monotonic function of the crowding parameter $\beta$ and thus the
existence of an optimal value, $\beta^{\rm opt}$, i.e. a value for which
$h$ attains its maximum. The latter depends on the nonlinearity bias
$\sigma$. In Fig.~\ref{fig:optbeta} we report results of some
dedicated simulations proving that the optimal value of the crowding
parameter is a decreasing function of the nonlinear bias $\sigma$, in
the case of uncorrelated scale free networks, for several values of
$\gamma$. A qualitatively similar result holds true also for other
network topologies (data not shown).

\begin{figure}[ht]  
\centering
\includegraphics[scale=0.28]{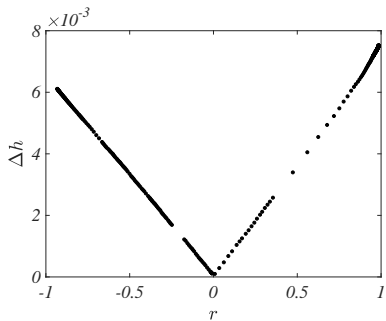}
\vspace{-1.5cm}
\caption{\textbf{Entropy rate for assortative/disassortative synthetic
    networks}.  We built several random assortative and disassortative synthetic networks, made of $N=1000$ nodes, each one characterised by its assortativity coefficient, $r\in [-1,1]$. We then compute the deviation between the entropy rate $h_{\rm HMF}$ and the analogous quantity obtained when degree correlations are silenced using a null model, $h_{\rm HMF}^{uncorr}$. We can observe a slight dependence of $\Delta h=\max_{\beta}| h/N - h_{\rm HMF}|$ on $r$, which in the worst case scenario reaches a few percents.}
\label{fig:Deltah}
\end{figure}

To study the impact of the assortativity on the entropy rate, we build $400$ random assortative and disassortative synthetic networks made of $N=1000$ nodes, each one characterised by its assortativity coefficient, $r\in [-1,1]$. For each network we computed the entropy rate using {Eq.~\eqref{eq:entropyrate2}} (in the main body of the paper), or equivalently Eq.~\eqref{eq:entropyrate9}, that is taking into account the possible correlations among nodes degree; then we applied a degree-preserving randomisation process to the network, namely we build a null model by rewiring links without changing the nodes degree. Eventually we computed the entropy rate by using the nodes degree distribution $P(k)$ common to both networks, via Eq.~\eqref{eq:entropyrate9}. This amounts in turn to neglect the degree correlations. Let us observe that in this way both networks exhibit the same asymptotic nodes densities which depend only on the node degree for fixed $\beta$ and $\sigma$, see Eq.~(4) (in the main body of the paper). Hence,  any possible differences as stemming from the usage of  Eq.~\eqref{eq:entropyrate9} and Eq.~(4) (in the main body of the paper) should be traced back to nodes degree correlations~\footnote{A randomisation algorithm preserving the total number of nodes and the average degree will generically induce different asymptotic nodes densities and thus deviation of $h_{\rm HMF}^{uncorr}$ from $h_{\rm HMF}$ could be imputed to both the nodes degree correlation and the different nodes densities.}. To measure the discrepancies as originated by the two aforementioned formulas, we define $\Delta h =\max_{\beta} |h_{\rm HMF}-h_{\rm HMF}^{uncorr}|$. The results reported in Fig.~\ref{fig:Deltah} show a dependence of the latter on $r$; $\Delta h$ vanishes for $r\rightarrow 0$ coherently with the fact that, for non-assortative networks $h_{\rm HMF}$ and $h_{\rm HMF}^{uncorr}$ should eventually coincide. Then, $\Delta h$ is maximal for $|r|\rightarrow 1$ and in the worst case scenario, the difference is bounded by a few percents.We also stress that this behaviour does not depend on the value of the nonlinear bias $\sigma$ imposed (data not shown).

\section{Real and synthetic Networks}
\label{ssec:nets}

The synthetic uncorrelated scale-free networks used in our analyses have
been created by means of the configuration model~\cite{LNR2017},
i.e. by drawing a set of positive integer numbers according to the
distribution $\sim 1/k^{\gamma}$, and then by using the latter as the
degree sequence. In the Erd\H os-{R\'enyi} networks with $N$ nodes, each
of the $N(N-1)/2$ edges is created independently with probability
$p_{\rm ER}$.

{To study the behaviour of our model of nonlinear
  random walkers on real-world networks we have selected various
  networks from different domains and with a different number of
  nodes, links and level of assortativity. The networks considered and
  the results obtained are summarized in Table~\ref{tab:table}.
In particular, we have evaluated the maximal entropy rate $h^{\rm
  opt}$ for each network, and we have compared this value to that
obtained in two different types of null models. The quantity  
$\Delta h_{\rm rew}^{\rm opt}$ (resp. $\Delta h_{\rm rnd}^{\rm opt}$) denotes 
the difference between the value of $h^{\rm
  opt}$ and that of a null model (both normalised to the system size)
that consists in randomising the original network 
under the assumption of preserving its  
node degrees (resp. the average degree) in the randomisation. In formulae:}
\begin{equation}
\label{eq:defdeltas}
\Delta h_{\rm rew}^{\rm opt}= \frac{h^{\rm opt} - h^{\rm opt}_{\rm null,rew}}{N}\quad\text{and}\quad \Delta h_{\rm rnd}^{\rm opt}= \frac{h^{\rm opt} - h^{\rm opt}_{\rm null,rnd}}{N}\, .
\end{equation} 

{Let us observe that, by definition, networks in the
  first type of null model have the same degree sequence as 
  the original real-world network. The asymptotic distribution in 
  Eq.~\eqref{eq:eqsol} depends, for a fixed load $\beta$, 
  only on the degree sequence of the network.  We can hence
  conclude that both the real network and the rewired
  ones exhibit the same asymptotic density of the walkers  
  $\rho_i^*$ for all $i=1,\dots, N$. 
  From the expression of the entropy rate in 
  Eq.~\eqref{eq:entropyrateSI} we can thus obtain that $h_{\rm null,rew}$
  differs from $h$ only for the following contribution coming from 
  differences in the adjacency matrices:}
%
\begin{equation*}
h-h_{\rm null,rew}=-\sum_{ij}\rho_i^{*} (a_{ij}-a_{ij}^{(\rm null,rew)})\frac{f(\rho_i^{*})g\left(1-\rho_j^{*}\right)}{k_i} \log \left[\frac{f(\rho_i^{*})g\left(1-\rho_j^{*}\right)}{k_i}\right]\, ,
\end{equation*}
%
{where $a_{ij}$ is the adjacency matrix of the
  original network while $a_{ij}^{(\rm null,rew)}$ the one obtained
  after the degree-preserving randomisation.
  Differences between the two matrices are limited by 
  be constraint imposed by fixing the degree sequence, and
  so are the differences between the entropy rates. From the values
  of  $\Delta h_{\rm rew}^{\rm opt}$ we observe that in general real-world
  networks,
  with the exception of Internet at the autonomous systems level and some
  social networks perform better in terms of explorability especially
in the case of walkers with large values of $\sigma$.}

\begin{figure}[ht]
\centering
\includegraphics[scale=0.21]{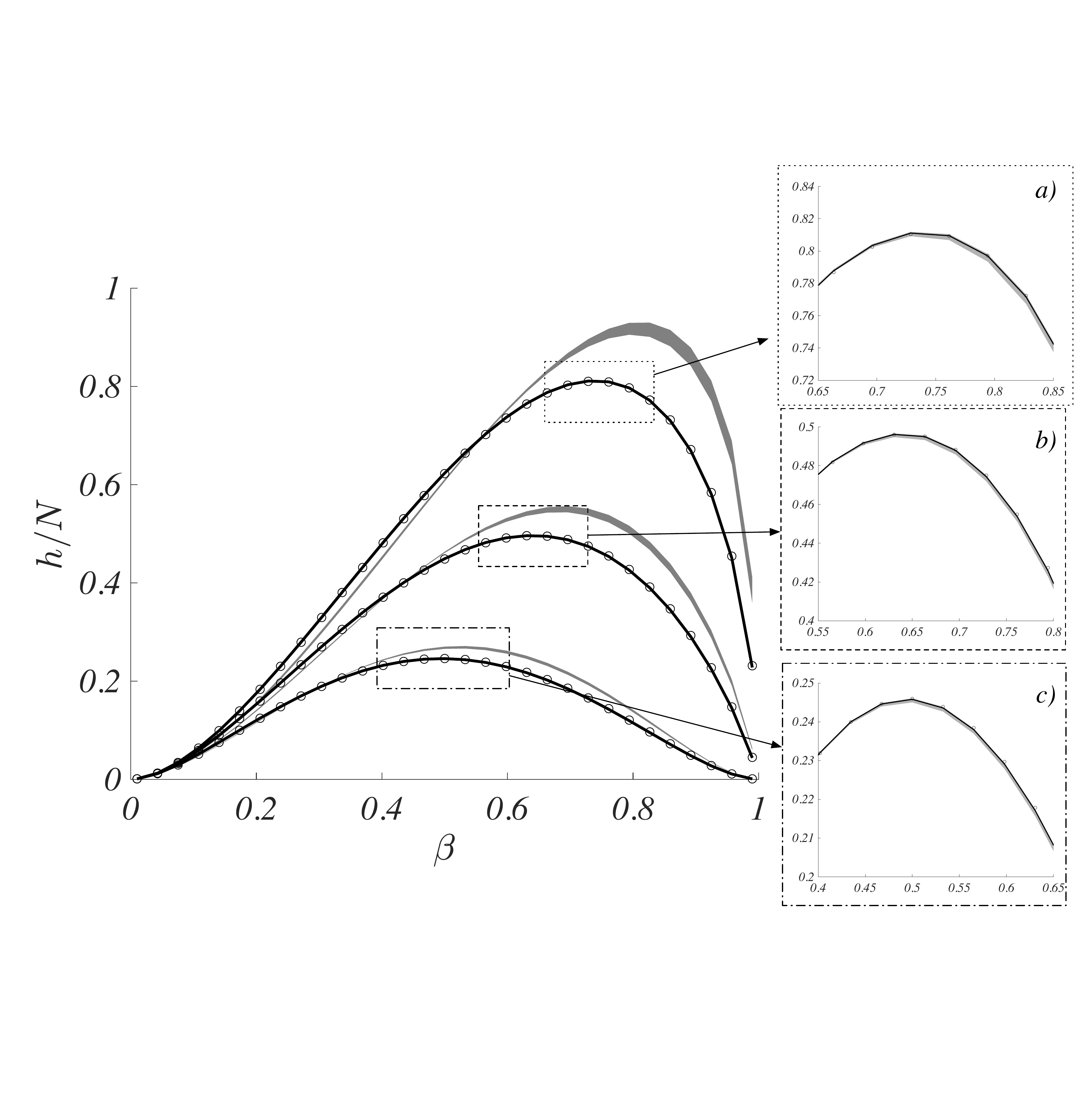}
\vspace{-1.5cm}
\caption{{\textbf{Comparison of the entropy rates for the different null models}. We report in the main panel the entropy rate for the real network ``C. Elegans frontal''~\cite{celeg} (circles), the one computed using the HMF assumption (black line) and the one for the randomised network preserving the average degree (shaded grey areas corresponding to the interval $[\min h/N,\max h/N]$ for the $50$ replicas). In the panels a-b-c we zoom in the vicinity of
    the maximum of the entropy rate to appreciate the similarity with the entropy rate obtained for the degree-preserving null model.}}
\label{fig:compareh}
\end{figure}

\begin{figure}[ht]
\centering
\includegraphics[scale=0.31]{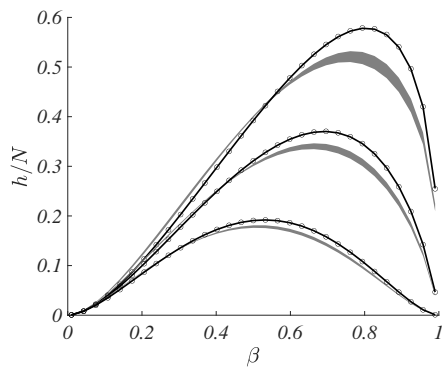}
\vspace{-1.5cm}
\caption{{\textbf{Entropy rates for the road networks}. We report the entropy rate for the road network of London~\cite{Crucitti2006,LNR2017} (circles) together with the same quantity computed using the randomised network preserving the average degree (shaded grey areas corresponding to the interval $[\min h/N,\max h/N]$ for the $50$ replicas). The solid black line denotes the entropy rate computed using the HMF assumption, that also coincide with the entropy rate obtained using the randomised model that preserves the degree distribution (not visible at this scale).}}
\label{fig:comparehroad}
\end{figure}

{On the other hand, using a null model 
  that only preserves the average degree of the original network,
  we will obtain larger
  variations of the entropy rates because now the asymptotic density
  will also differ (see Fig.~\ref{fig:compareh}). Randomised
  networks obtained in this way in
  general exhibit larger maximal entropy rates, with the notable exception of
  urban street patterns (see Table~\ref{tab:table} and
  Fig.~\ref{fig:comparehroad}) that have not only a positive
  value of 
   $\Delta h_{\rm rew}^{\rm opt}$, but also always a positive value of $\Delta h_{\rm rnd}^{\rm opt}$. In these systems, in fact, 
  randomised networks with the same average degree performs
  worse in term of explorability, namely their entropy rate is always
  smaller than the one of the original network. Based on
  this it is tempting to speculate that road network have been
  assembled so as to optimise their structure for transport under congested
  conditions: any randomised version,
  that disrupt the local organisation of crossroads, will perform
  worse.}
  
  {Our results imply that the entropy rate per node, for the synthetic networks, the real ones or a randomised version of these latter, exhibits the same functional behaviour; $h$ vanishes for very small and very large values of $\beta$ and then achieves a single maximum at an intermediate value of the parameter. This value, $\beta^{\rm opt}$, corresponds thus to an optimal network crowding (the total load on the network) that facilitates the exploration of the network, being a maximum of the entropy rate. In Table~\ref{tab:tablenew} we report, for three choices of the nonlinear parameter $\sigma$, the values of the optimal crowding computed for the real networks presented in Table~\ref{tab:table}, together with the corresponding differences with respect to the same quantities computed for the null model networks obtained by randomising the original network  under the assumption of preserving its node degrees (resp. the average degree). In formulae:
\begin{equation}
\label{eq:defdeltasbeta}
\Delta \beta_{\rm rew}^{\rm opt}= \beta^{\rm opt} - \beta^{\rm opt}_{\rm null,rew}\quad\text{and}\quad \Delta \beta_{\rm rnd}^{\rm opt}= \beta^{\rm opt} - \beta^{\rm opt}_{\rm null,rnd}\, .
\end{equation} 
}

\begin{figure*}[ht]
\centering
 \includegraphics[scale=0.21]{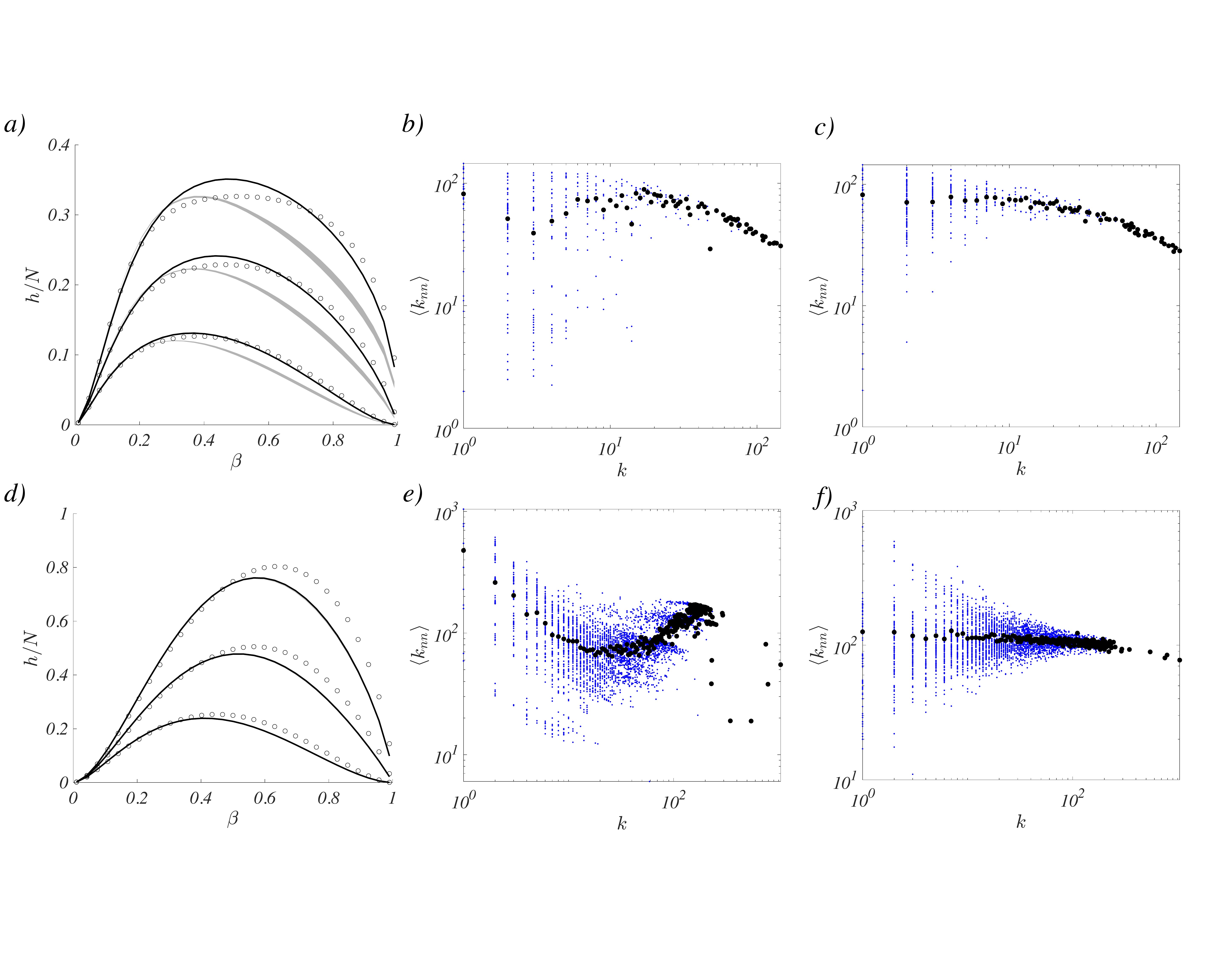}
\vspace{-1.5cm}
\caption{{\textbf{The role of degree-degree 
      correlations}. Results reported for the US Airplane
    network~\cite{Colizza2007,LNR2017} (top panels) and
    Facebook~\cite{facebook} (bottom panels) indicate that, when
    the randomisation (preserving the degree distribution) is not
    able to completely wash out the degree-degree correlations,
    then the entropy rate of the randomised network 
    (grey shaded area in panels a and d) differs from the one
    computed in the HMF approximation (black lines in panels a
    and d). In both cases, the circles denote the entropy rate for the
    original network. We report for the original network (panels b and
    e ) and the randomised one (panels c and f) a measure of the presence of
    degree-degree correlations, namely the average degree
    $\langle k_{nn}\rangle$ of neighbours of nodes of degree $k$ as a
    function of $k$. The similarity of the scatter
    plots in panels b and c suggest the presence of degree-degree
    correlations in the randomised version of the US airplane network.
    Comparison of panels e and
    f suggests instead that correlations have been destroyed in the 
    case of Facebook.}}
\label{fig:comparehcorrel}
\end{figure*}

\medskip
{To conclude, let us consider the role of the degree-degree 
  correlations on the computation of the entropy rate of nonlinear random walkers. In few cases,
  e.g.  the three Internet Autonomous Systems networks and the US
  Airplane network the randomisation process is not able to completely
  destroy the degree correlations and thus the entropy rate for the
  null model will deviated from the same quantity computed in the 
  HMF approximation (see Fig.~\ref{fig:comparehcorrel} top panels for
  the case of the US Airplane network). On the other hand, once the
  randomisation is able to wash out the degree-degree correlations, we obtain a
  satisfying agreement between the null model entropy rate and the HMF
  approximation. We recall in fact that the HMF approximation implies neglecting
  degree-degree correlations (see Fig.~\ref{fig:comparehcorrel} bottom
  panels for the case of the Facebook network).}

\newpage
\begin{sidewaystable}
\begin{tabular}{lc|c|c|c|c|c|c|}   
  \hline
\hspace{1em}&  &   &  & $\sigma=0.5$  & $\sigma=1$ & $\sigma=2$ \\
\hspace{1em} {\small Networks} & {\small $N$} & $\langle k \rangle$ & $r$ & $\Delta h_{\rm rew}^{\rm opt} $ , $\Delta h_{\rm rnd}^{\rm opt}$  & $\Delta h_{\rm rew}^{\rm opt} $ , $\Delta h_{\rm rnd}^{\rm opt}$ & $\Delta h_{\rm rew}^{\rm opt} $ , $\Delta h_{\rm rnd}^{\rm opt}$ \\
\hline
\hspace{1em}{\small Facebook}~\cite{facebook} & $4039$ & $43.69$ & $0.0636$ & ${0.5\, 10^{-1}}$ , $-6.0\,10^{-1}$& ${0.3\, 10^{-1}}$ , $-3.0\,10^{-1}$& $0.2\, 10^{-1}$ ,$-1.2\,10^{-1}$ \\
\hspace{1em}{\small US Airplane}~\cite{Colizza2007,LNR2017}& $500$ & $11.92$ & $-0.2679$ & ${0.3\, 10^{-3}}\,^*$ , $-6.0\,10^{-1}$& ${0.6\, 10^{-2}}$ , $-3.3\,10^{-1}$& $0.6\, 10^{-2}$ , $-1.5\, 10^{-1}$\\  
\hspace{1em}{\small Internet Autonomous Systems network}&   &   &   &  &  &  \\  
\hspace{1em}{\hspace{2cm}\small (AS-19971108)}~\cite{LNR2017}& $3015$ & $3.42$ & $-0.2289$ & ${-1.8\, 10^{-2}}$ , $-2.9\,10^{-1}$& ${-1.3\, 10^{-2}}$ , $-1.7\,10^{-1}$& $-7.8\, 10^{-3}$ , $-8.3\,10^{-2}$\\  
\hspace{1em}{\small Internet Autonomous Systems network}&   &   &   &  &  &  \\  
\hspace{1em}{\hspace{2cm}\small (AS-19980402)}~\cite{LNR2017}& $3522$ & $3.59$ & $-0.210$ & ${-1.9\, 10^{-2}}$ , $-3.0\,10^{-1}$& ${-1.4\, 10^{-2}}$ , $-1.8\,10^{-1}$& $-8.0\, 10^{-3}$ , $-8.5\,10^{-2}$\\  
\hspace{1em}{\small Internet Autonomous Systems network}&   &   &   &  &  &  \\  
\hspace{1em}{\hspace{2cm}\small (AS-19980703)}~\cite{LNR2017}& $3797$ & $3.65$ & $-0.207$ & ${-1.8\, 10^{-2}}$ , $-3.0\,10^{-1}$& ${-1.3\, 10^{-2}}$ , $-1.8\,10^{-1}$& $-7.8\, 10^{-3}$ , $-8.5\,10^{-2}$\\  
\hspace{1em}{\small C. Elegans frontal}~\cite{celeg} & $131$ & $10.49$ & $0.0218$ & $1.2\, 10^{-5}\,^*$ , $-1.1\,10^{-1}$ & $4.5\, 10^{-4}$ , $-5.4\,10^{-2}$& $4.6\, 10^{-4}$ , $-2.3\,10^{-2}$\\  
\hspace{1em}{\small Ahmedabad}~\cite{Crucitti2006,LNR2017} & $2870$ & $3.05$ & $0.025$ & $0.3\, 10^{-3}$ , $8.1\,10^{-2}$ & ${-0.2\, 10^{-5}}\,^*$ , $4.2\,10^{-2}$ & $0.3\, 10^{-5}\,^*$ , $1.9\,10^{-2}$\\  
\hspace{1em}{\small Barcelona}~\cite{Crucitti2006,LNR2017} & $210$ & $3.08$ & $-0.037$ & $1.0\, 10^{-3}$ , $3.2\,10^{-2}$& $0.5\, 10^{-3}$ , $1.6\,10^{-2}$& $0.1\, 10^{-3}$ , $6.7\,10^{-3}$\\  
\hspace{1em}{\small Bologna}~\cite{Crucitti2006,LNR2017}& $541$ & $2.85$ & $0.024$ & $0.3\, 10^{-3}$ , $5.2\,10^{-2}$ & $0.2\, 10^{-3}$ , $2.7\,10^{-2}$& $0.9\, 10^{-4}$ , $1.2\,10^{-2}$\\  
\hspace{1em}{\small Cairo}~\cite{Crucitti2006,LNR2017} & $1496$ & $3.01$ & $-0.0057$ & $0.1\, 10^{-3}$ , $7.8\,10^{-2}$& $0.6\, 10^{-4}$ , $4.1\,10^{-2}$& $0.1\, 10^{-4}$ , $1.8\,10^{-2}$\\  
\hspace{1em}{\small London}~\cite{Crucitti2006,LNR2017} & $488$ & $2.99$ & $-0.0301$ & $0.6\, 10^{-3}$ , $5.7\,10^{-2}$& $0.3\, 10^{-3}$ , $3.0\,10^{-2}$& $0.7\, 10^{-4}$ , $1.3\,10^{-2}$\\  
\hspace{1em}{\small Venice}~\cite{Crucitti2006,LNR2017} & $1840$ & $2.61$ & $-0.0921$ & $1.4\, 10^{-3}$ , $3.8\, 10^{-2}$ & $0.6\, 10^{-3}$ , $1.9\,10^{-2}$ & $0.1\, 10^{-3}$ , $8.6\, 10^{-3}$\\  
\hspace{1em}{\small Medicis Family}~\cite{Wasserman1994,LNR2017} & $15$ & $2.67$ & $-0.3748$ & ${-0.8\, 10^{-3}}\,^*$ , $-8.2\, 10^{-3}$& ${-0.7\, 10^{-3}}\,^*$ , $-4.6\,10^{-3}$& ${-0.4\, 10^{-3}}\,^*$ , $-1.9\,10^{-3}$\\  
\hspace{1em}{\small Zachary's Karate Club}~\cite{LNR2017} & $34$ & $4.59$ & $-0.4756$ & ${-1.5\, 10^{-3}}$ , $-1.2\,10^{-1}$& ${-2.0\, 10^{-3}}$ , $ -6.3\,10^{-2}$& ${-1.6\, 10^{-3}}$ , $-2.7\,10^{-2}$\\  
\hspace{1em}{\small Kindergarten}~\cite{LNR2017} & $16$ & $4.88$ & $0.2234$ & ${-1.2\, 10^{-3}}$ , $-2.1\,10^{-2}$ & $0.4\, 10^{-3}$ , $-1.2\, 10^{-2}$ & $0.7\, 10^{-3}$ , $-5.2\, 10^{-3}$\\  
\hspace{1em}{\small Primates}~\cite{Everett1999,LNR2017} & $14$ & $4.43$ & $-0.5046$ & ${-0.6\, 10^{-3}}$ , $-1.0\, 10 ^{-1}$ & ${-0.1\, 10^{-3}}\,^*$, $-5.2\, 10^{-2}$ & ${0.7\, 10^{-4}}\,^*$, $-2.3\, 10^{-2}$\\  
\end{tabular}
\caption{{\textbf{Nonlinear random walkers on
      real-world networks}.
    Number of nodes $N$, average node degree $\langle k\rangle$ and
    degree-degree correlation coefficient  $r$~\cite{Newman2002,Newman2003} 
    for each of the networks considered. For three
    different choices of the nonlinear bias, namely $\sigma=0.5, 1$ and 2,
    we report the difference between the maximum value
    of the entropy rate $h^{\rm opt}$ in the original network and in
    our two null models, $\Delta h_{\rm rew}^{\rm opt}$ and
    $\Delta h_{\rm rnd}^{\rm opt}$. $50$ different network 
    realisations have been considered to evaluate averages and
    standard deviations for each null model. 
    Values marked by $*$ are those smaller than one standard
    deviation.}}
\label{tab:table}
\end{sidewaystable}

\newpage
\begin{sidewaystable}
{
\begin{tabular}{lc|c|c|c|c|c|c|}   
  \hline
\hspace{1em}& $\sigma=0.5$  & $\sigma=1$ & $\sigma=2$ \\
\hspace{1em} {\small Networks} & $\beta^{\rm opt}$ , $\Delta \beta_{\rm rew}^{\rm opt} $ , $\Delta \beta_{\rm rnd}^{\rm opt}$  & $\beta^{\rm opt}$ , $\Delta \beta_{\rm rew}^{\rm opt} $ , $\Delta \beta_{\rm rnd}^{\rm opt}$ & $\beta^{\rm opt}$ , $\Delta \beta_{\rm rew}^{\rm opt} $ , $\Delta \beta_{\rm rnd}^{\rm opt}$ \\
\hline
\hspace{1em}{\small Facebook}~\cite{facebook} 
& ${6.3\, 10^{-1}}$, ${6.1\, 10^{-2}}$ , $-2.3\,10^{-1}$ 
& ${5.7\, 10^{-1}}$, ${5.1\, 10^{-2}}$ , $-1.7\,10^{-1}$
& $4.5\, 10^{-1}$ ,$4.0\, 10^{-2}$ ,$-1.0\,10^{-1}$ \\
\hspace{1em}{\small US Airplane}~\cite{Colizza2007,LNR2017}
& ${5.1\, 10^{-1}}$, ${1.2\, 10^{-1}}$ , $-3.2\,10^{-1}$ 
& ${4.6\, 10^{-1}}$, ${9.2\, 10^{-2}}$ , $-2.4\,10^{-1}$
& $3.8\, 10^{-1}$ ,$6.7\, 10^{-2}$ ,$-1.6\,10^{-1}$ \\
\hspace{1em}{\small Internet Autonomous Systems network}&   &   &  \\  
\hspace{1em}{\hspace{2cm}\small (AS-19971108)}~\cite{LNR2017}
& ${6.3\, 10^{-1}}$, ${-7.6\, 10^{-2}}$ , $-3.1\,10^{-2}$ 
& ${5.3\, 10^{-1}}$, ${-6.7\, 10^{-2}}$ , $-3.5\,10^{-2}$
& $4.1\, 10^{-1}$ ,$-5.0\, 10^{-2}$ ,$-3.1\,10^{-2}$ \\
\hspace{1em}{\small Internet Autonomous Systems network}&   &   & \\  
\hspace{1em}{\hspace{2cm}\small (AS-19980402)}~\cite{LNR2017}
& ${6.3\, 10^{-1}}$, ${-7.2\, 10^{-2}}$ , $-3.5\,10^{-2}$ 
& ${5.4\, 10^{-1}}$, ${-6.4\, 10^{-2}}$ , $-3.7\,10^{-2}$
& $4.1\, 10^{-1}$ ,$-4.8\, 10^{-2}$ ,$-3.3\,10^{-2}$ \\
\hspace{1em}{\small Internet Autonomous Systems network}&   &   & \\  
\hspace{1em}{\hspace{2cm}\small (AS-19980703)}~\cite{LNR2017}
& ${6.3\, 10^{-1}}$, ${-7.0\, 10^{-2}}$ , $-3.3\,10^{-2}$ 
& ${5.4\, 10^{-1}}$, ${-6.2\, 10^{-2}}$ , $-3.6\,10^{-2}$
& $4.2\, 10^{-1}$ ,$-4.7\, 10^{-2}$ ,$-3.2\,10^{-2}$ \\
\hspace{1em}{\small C. Elegans frontal}~\cite{celeg}
& ${7.3\, 10^{-1}}$, ${2.2\, 10^{-3}}$ , $-6.3\,10^{-2}$ 
& ${6.4\, 10^{-1}}$, ${2.3\, 10^{-3}}$ , $-4.4\,10^{-2}$
& $5.0\, 10^{-1}$ ,$2.1\, 10^{-3}$ ,$-2.8\,10^{-2}$ \\
\hspace{1em}{\small Ahmedabad}~\cite{Crucitti2006,LNR2017}
& ${8.3\, 10^{-1}}$, ${3.1\, 10^{-3}}$ , $7.0\,10^{-2}$ 
& ${7.1\, 10^{-1}}$, ${1.7\, 10^{-3}}$ , $4.8\,10^{-2}$
& $5.4\, 10^{-1}$ ,$4.2\, 10^{-4}$ ,$2.9\,10^{-2}$ \\
\hspace{1em}{\small Barcelona}~\cite{Crucitti2006,LNR2017}
& ${7.7\, 10^{-1}}$, ${5.7\, 10^{-4}}$ , $9.8\,10^{-3}$ 
& ${6.7\, 10^{-1}}$, ${-5.4\, 10^{-5}}$ , $7.7\,10^{-3}$
& $5.2\, 10^{-1}$ ,$-4.9\, 10^{-4}$ ,$5.5\,10^{-3}$ \\
\hspace{1em}{\small Bologna}~\cite{Crucitti2006,LNR2017}
& ${7.9\, 10^{-1}}$, ${4.0\, 10^{-4}}$ , $3.6\,10^{-2}$ 
& ${6.9\, 10^{-1}}$, ${2.0\, 10^{-4}}$ , $2.5\,10^{-2}$
& $5.3\, 10^{-1}$ ,$8.0\, 10^{-6}$ ,$1.6\,10^{-2}$ \\
\hspace{1em}{\small Cairo}~\cite{Crucitti2006,LNR2017} 
& ${8.2\, 10^{-1}}$, ${4.6\, 10^{-5}}$ , $6.6\,10^{-2}$ 
& ${7.0\, 10^{-1}}$, ${-8.7\, 10^{-6}}$ , $4.5\,10^{-2}$
& $5.4\, 10^{-1}$ ,$-5.1\, 10^{-5}$ ,$2.8\,10^{-2}$ \\
\hspace{1em}{\small London}~\cite{Crucitti2006,LNR2017}
& ${8.0\, 10^{-1}}$, ${2.7\, 10^{-4}}$ , $4.3\,10^{-2}$ 
& ${6.9\, 10^{-1}}$, ${-9.5\, 10^{-5}}$ , $2.9\,10^{-2}$
& $5.3\, 10^{-1}$ ,$-3.3\, 10^{-4}$ , $1.8\,10^{-2}$ \\
\hspace{1em}{\small Venice}~\cite{Crucitti2006,LNR2017}
& ${7.8\, 10^{-1}}$, ${5.3\, 10^{-4}}$ , $2.0\,10^{-2}$ 
& ${6.8\, 10^{-1}}$, ${-4.1\, 10^{-4}}$ , $1.5\,10^{-2}$
& $5.2\, 10^{-1}$ ,$-9.1\, 10^{-4}$ , $9.9\,10^{-3}$ \\
\hspace{1em}{\small Medicis Family}~\cite{Wasserman1994,LNR2017}
& ${7.6\, 10^{-1}}$, ${-2.6\, 10^{-3}}$ , $-1.2\,10^{-2}$ 
& ${6.6\, 10^{-1}}$, ${-1.8\, 10^{-3}}$ , $-6.4\,10^{-3}$
& $5.1\, 10^{-1}$ ,$-1.3\, 10^{-3}$ , $-2.9\,10^{-3}$ \\
\hspace{1em}{\small Zachary's Karate Club}~\cite{LNR2017}
& ${7.1\, 10^{-1}}$, ${-9.8\, 10^{-3}}$ , $-3.9\,10^{-2}$ 
& ${6.2\, 10^{-1}}$, ${-9.5\, 10^{-3}}$ , $-3.3\,10^{-2}$
& $4.8\, 10^{-1}$ ,$-8.1\, 10^{-3}$ , $-2.2\,10^{-2}$ \\
\hspace{1em}{\small Kindergarten}~\cite{LNR2017}
& ${7.7\, 10^{-1}}$, ${3.9\, 10^{-3}}$ , $-2.6\,10^{-2}$ 
& ${6.7\, 10^{-1}}$, ${4.1\, 10^{-3}}$ , $-1.7\,10^{-2}$
& $5.2\, 10^{-1}$ ,$3.6\, 10^{-3}$ , $-1.0\,10^{-2}$ \\
\hspace{1em}{\small Primates}~\cite{Everett1999,LNR2017}
& ${7.3\, 10^{-1}}$, ${8.6\, 10^{-4}}$ , $-4.9\,10^{-2}$ 
& ${6.3\, 10^{-1}}$, ${1.1\, 10^{-3}}$ , $-3.7\,10^{-2}$
& $4.9\, 10^{-1}$ ,$1.1\, 10^{-3}$ , $-2.4\,10^{-2}$ \\
\end{tabular}
\caption{{\textbf{Nonlinear random walkers on
      real-world networks (II)}.  For three
    different choices of the nonlinear bias, namely $\sigma=0.5, 1$ and 2,
    we report the optimal value of the crowding parameter, $\beta^{\rm opt}$, for the original network and 
    and in the two null models we used (degree distribution and average degree preserving) $\Delta \beta^{\rm opt}_{rew}$ and $\Delta \beta^{\rm opt}_{rnd}$. $50$ different network 
    realisations have been considered to evaluate averages and
    standard deviations for each null model.}}
\label{tab:tablenew}}
\end{sidewaystable}

\newpage

\section{Different choices of the bias functions $f$ and $g$}
\label{ssec:morebias}

\begin{figure}[ht]
\centering
\includegraphics[scale=0.26]{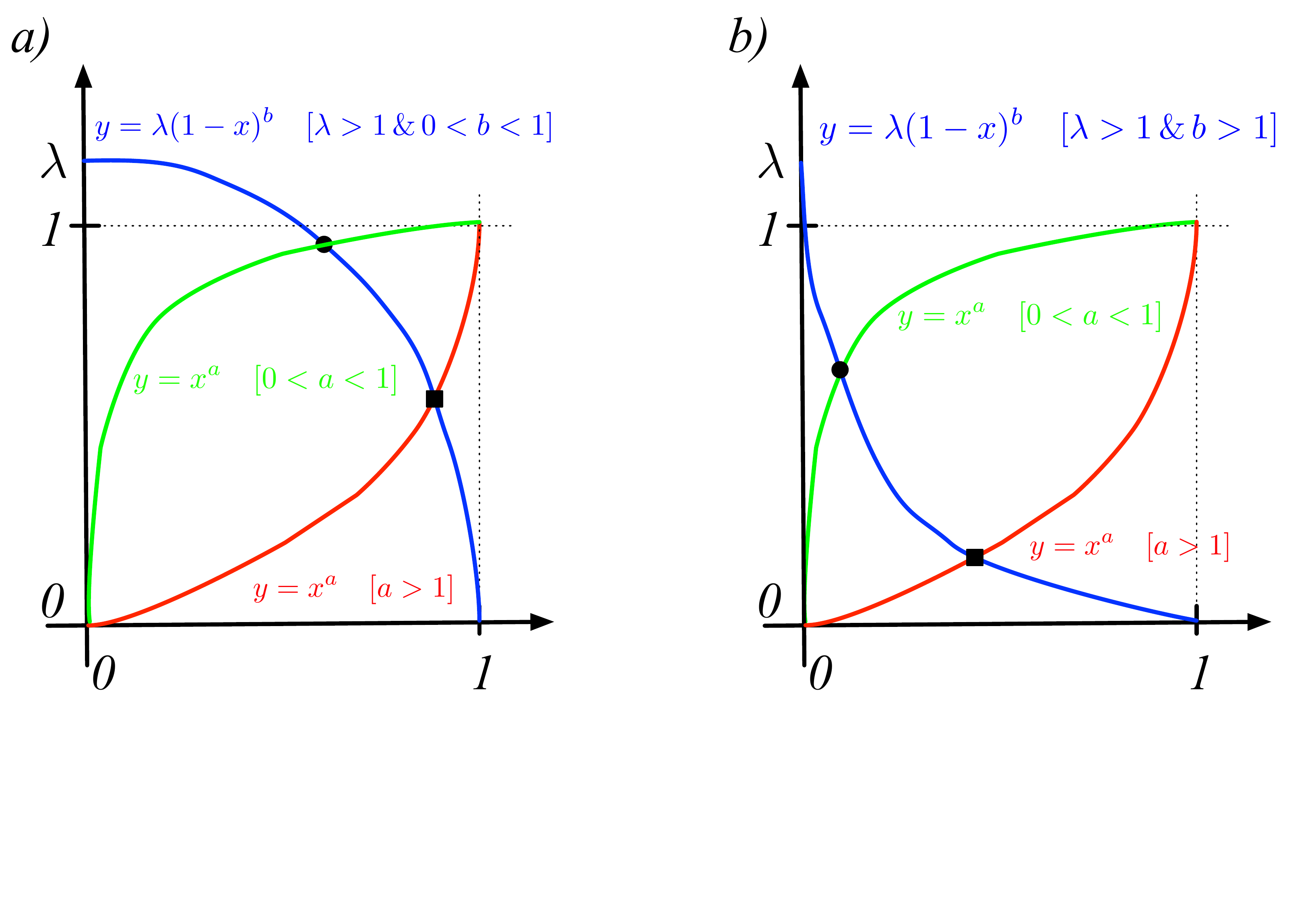}
\vspace{-1cm}
\caption{Proof of the existence and uniqueness of the stationary solution via a qualitative approach. The solution of Eq.~\eqref{eq:statsol4} can be obtained as the intersection of the curves $y=x^{\alpha}$ (red for $\alpha
  >1$ and green for $\alpha<1$) and $y=\lambda (1-x)^{\sigma}$ (blue), where $\lambda = ck_i$ and $x=\rho^*_i$. For any choice of $\alpha>0$, $\sigma>0$ and $\lambda>0$ such two curves admit one and only one intersection in $[0,1]$; left panel $0<\sigma<1$, right panel: $1<\sigma$.}
\label{fig:intersection}
\end{figure}

The aim of this section is to introduce and study some more general classes of biases functions $f$ and $g$. The simplest straightforward generalisation is to consider
\begin{eqnarray*}
f(x)&=&x^{\alpha}\text{ and } \\ g(x)&=& (1-x)^{\sigma} \text{for $0 \leq  x< 1$ and $0$ otherwise}\, ,
\end{eqnarray*}
for some real positive $\alpha$ and $\sigma$. Under such assumption the asymptotic solution given by Eq. (4) (in the main body of the paper) rewrites:
\begin{equation}
\frac{(\rho_i^{*})^{\alpha}}{\left(1 - \rho_i^{*}\right)^{\sigma}}=c_{\sigma}k_i\quad \forall i \, .
\label{eq:statsol4}
\end{equation}
Observe that the function on the left hand side equals $0$ for $\rho_i^*=0$ and monotonically diverge to $+\infty$ for $\rho_i^{*}\rightarrow 1^{-}$. Hence, for any $c_{\sigma}k_i$, there exists one and only one value for $\rho_i^*$ satisfying the equality. Stated differently looking for the asymptotic solution corresponds to finding the intersection of the two curves $y=x^{\alpha}$ and $y=\lambda (1-x)^{\sigma}$ (see Fig.~\ref{fig:intersection} for two generic cases), and by continuity such intersection is alway unique.

\end{document}